\documentclass[iop]{emulateapj}

\usepackage{amsmath,amssymb,times,graphics,morefloats,tipa,color}
\bibstyle{apj}

\newcommand{\lha}{\ensuremath{{\rm L}_{\rm H\alpha}/{\rm L}_{\rm bol}}}
\newcommand{\rz}{\ensuremath{r-z}}
\newcommand{\uz}{\ensuremath{u-z}}
\newcommand{\fnuv}{\ensuremath{fuv-nuv}}
\newcommand{\JH}{\ensuremath{J-H}}

\defcitealias{W11}{W11}
\defcitealias{W08}{W08}

\begin{document}

\title{The Effects of Close Companions (and Rotation) on the Magnetic Activity of M
  Dwarfs}
\journalinfo{{\it The Astronomical Journal}, submitted}
\submitted{Submitted: May 10, 2012; Accepted: June 28, 2012}

\author{Dylan P. Morgan\altaffilmark{1,2}, Andrew A. West\altaffilmark{1}, Ane Garc\'{e}s\altaffilmark{3},
  Silvia Catal\'{a}n\altaffilmark{4}, Saurav Dhital\altaffilmark{5,1}, Miriam Fuchs\altaffilmark{6,1},and Nicole M. Silvestri\altaffilmark{7}
}

\altaffiltext{1}{Astronomy Department, Boston University, 725 Commonwealth
  Ave, Boston, MA 02215, USA}
\altaffiltext{2}{Corresponding author: dpmorg@bu.edu}
\altaffiltext{3}{Institut de Ci\`encies de l'Espai (IEEC-CSIC), Facultat de Ci\`encies, Campus UAB, 08193 Bellaterra, Spain}
\altaffiltext{4}{Centre for Astrophysics Research, University of
  Hertfordshire, College Lane, Hatfield, AL10 AB, UK}
\altaffiltext{5}{Department of Physics \& Astronomy, Vanderbilt University, 6301 Stevenson Center, Nashville, TN 37235, USA}
\altaffiltext{6}{Department of Physics and Astronomy, Haverford
  College, 370 Lancaster Avenue,
  Haverford, PA 19041, USA}
\altaffiltext{7}{Department of Astronomy, University of Washington, Box
  351580, Seattle, WA 98195, USA}

\begin{abstract}
We present a study of close white dwarf and M dwarf (WD$+$dM) binary
systems and examine the effect that a close companion has on the
magnetic field generation in M dwarfs.  We use a base sample of 1602 white
dwarf -- main sequence binaries from \cite{Rebassa2010} to develop a set of color cuts in GALEX, SDSS, UKIDSS, and 2MASS
color space.  Then using the SDSS DR8 spectroscopic database, we
construct a sample of 1756 WD$+$dM high-quality pairs from our color cuts and previous catalogs.  We
separate the individual WD and dM from each spectrum using an iterative
technique that compares the WD and dM components to best-fit
templates.  Using the absolute height above the Galactic plane as a
proxy for age, and the H$\alpha$ emission line as an indicator for
magnetic activity, we investigate the age-activity relation for our
sample for spectral types $\leq$ M7.  Our results show that early-type
M dwarfs ($\leq$ M4) in close binary systems are more likely to be active and have longer activity lifetimes compared to their field counterparts. However, at a spectral type of
M5 (just past the onset of full convection in M dwarfs), the
activity fraction and lifetimes of WD$+$dM binary systems becomes more comparable
to that of the field M dwarfs.  One of the implications of having a close
binary companion is presumed to be increased stellar rotation through
disk-disruption, tidal effects, or angular momentum exchange.  Thus,
we interpet the similarity in activity behavior between late-type dMs in
WD$+$dM pairs and late-type field dMs to be due to a decrease in
sensitivity in
close binary companions (or stellar rotation), which has implications
for the nature of magnetic activity in fully-convective stars.  Using
the WD components of the pairs, we find WD cooling ages to use as an
additional constraint on the age--activity relation for our sample.  We
find that, on average, active early-type dMs tend to be younger and
that active late-type dMs span a much broader age regime making them
indistinguishable from the inactive late-type population.  We also show that magnetic strength, as
measured by H$\alpha$, is comparable between paired and field M dwarfs until
a spectral type of M6/M7 where M dwarf activity for stars with close companions becomes much
stronger.  In addition, we present 37 very-close candidate pairs with
fast-moving orbits that display radial velocity changes over hour time-scales.
\end{abstract}


\section{Introduction}\label{Sec:Intro}
Aside from being cooler, dimmer, and much less massive than the stars
like the Sun, M dwarfs (dM) are the major stellar constituents 
of the Galaxy.  The ubiquitous nature of dMs, in addition to their long 
lifetimes, make them excellent tracers of Galactic structure,
kinematics, and evolution \citep{Gizis2002, W04}.  In addition, dMs have strong magnetic fields that heat their upper atmospheres, producing
excess emission in the chromosphere and corona (called ``magnetic
activity'').  This magnetic activity plays a vital role in the space
weather environments of dMs and likely affects the true habitability
of attending planets. How these small stars generate such strong
fields ($\sim$several kG) and the subsequent magnetic activity is
still not well understood.  Therefore, detailed observations of
the magnetic properties of dMs are vital for constraining theoretical dynamo
models.  Adding to their utility, the dM sequence (M0-M9) also probes an important
transition in the structure of stellar interiors, where they change from having both
radiative and convective zones, to being fully convective \citep[around a spectral type of M4; e.g.,][]{Dantona1985,Dorman1989,ChabrierBaraffe1997}.
This transition has important ramifications for how magnetic fields are generated and
maintained across the dM stellar sequence \citep[e.g.,][]{Morin2008,Reiners2009,Morin2010}.

 In the Sun, magnetic field generation is
thought to be primarily driven by the shear between the stratified 
solid body rotation of the radiative zone and the differential
rotation of the convective zone
\citep{Parker1955,Parker1993,Ossen2003}.  This transition region, called
the tachocline, acts to strengthen and stretch
poloidal magnetic fields, establishing a large-scale toroidal magnetic
field that can become unstable and rise through the convective zone to
become an active region on the surface of the star
\citep[e.g.,][]{Browning2008}.  Because of the importance of the
tachocline shearing in this model, stellar rotation plays a major
role in the strength of the magnetic field and resulting activity.  Magnetic field generation in
early-type dMs ($\leq$ M4) is also thought to be due to a similar
mechanism as early-type dMs still have
tachoclines.  However, dMs later than $\sim$M4 (late-type) are modeled to be fully
convective \citep[e.g.,][]{ChabrierBaraffe1997} and lack
tachoclines.  Despite these structural differences, many studies
have shown that late-type dMs have strong
fields and are
magnetically active (the fraction of stars that are active peaks around M7;~\citealt{Hawley1996};~\citealt{Gizis2000};~\citealt{W04,W06,W08}, hereafter W08;~\citealt{Donati08};~\citealt{Reiners2007,Reiners2009,Reiners2010})

In most stars, the strength (and presence) of magnetic activity
appears to correlate with age and rotation rate (as suggested
above).  Several decades ago,
\cite{WilsonWoolley1970} discovered a relation linking magnetic activity (emission in Ca II H\&K) and
the age of a star (estimated from galactic orbital parameters) using
main sequence stars (mostly G dwarfs). They found that stars that had undergone many dynamical interactions (i.e., are older) had
 lower levels of activity than those stars that had more circular orbits (i.e., are younger).  A short time later, \cite{Skumanich1972} discovered an age--rotation--activity relation (again using
 mostly G dwarfs) showing that stars
 tend to spin down as they age as well as decrease in activity. The
 Skumanich relation occurs because angular
momentum is lost from magnetized stellar winds; as a result, magnetic
fields (and activity) are reduced in magnitude.
 Subsequent studies have confirmed the \cite{Skumanich1972} results
 and 
 established relations between age and activity (e.g.,~\citealt{Wielen1977};~\citealt{Giampapa1986};~\citealt{Soderblom1991};~\citealt{Hawley1996,Hawley1999,Hawley2000};~\citealt{W04,W06,W08};~\citealt{MamajekHillenbrand2008}) as well
 as age and rotation \citep[e.g.,][]{Barry1988,Soderblom1991,MohantyBasri2003,Barnes2007,Kiraga2007,MamajekHillenbrand2008,Collier2009,Barnes2010}
 for G, K and early-type M dwarfs.

The exact mechanism behind the magnetic
field generation in late-type dMs is still unclear.  A recent
study \citep{Chabrier2006} suggested that magnetic activity in
late-type dMs is due to an $\alpha^{2}$ mechanism, where flux tubes are twisted in a
stratified rotating fluid acted upon by the Coriolis force.
\cite{Browning2008} used simulations to show that strong magnetic
fields are generated in fully convective envelopes when large slowly overturning
flows are acted upon by Coriolis forces.  It appears that no matter
the mechanism, 
stellar rotation still plays an important role; rotation has a strong influence on the magnitude of the
Coriolis force acting upon the overturning flows.  

There is some observational evidence that the age--activity and rotation--activity
relations extends into the late-type dM regime (\citealt{Delfosse1998,MohantyBasri2003,Reiners2007,Reiners2008}; W08).  However,  there are some clear difference between early and late-type dMs.  W08 calibrated a ``Galactic stratigraphy'' technique (stars farther from the Galactic mid-plane are statistically older) to show that activity decreases as a function of age in all dM spectral types and that the activity lifetimes of M dwarfs change across the transition to fully convective interiors (from $\sim$1-2 Gyr to $\sim$7-8 Gyr).  In addition,  \cite{Reiners2008} demonstrated that the vast majority of late-type dMs are rotating quickly (as compared to a distribution of rotation rates for early-type dwarfs).  These results suggest that if rotation is a strong driver of magnetic activity, then the spindown times of late-type dMs are much longer than those of more massive stars.   In a few cases, there have been reports of fast rotating dMs that do not show measurable magnetic activity \citep{WestBasri2009,Reiners2012}.  While some of these results may be due to spurious measurements \citep{Reiners2012}, they leave open the question of exactly what role rotation plays in the generation of magnetic field in fully convective stars.

M dwarfs with close binary companions may shed some light on the role that rotation plays in the generation, strength and lifetime of magnetic activity in dMs.  It is thought that the rotational evolution of isolated stars is intricately tied to the lifetime of the circumstellar disk \citep[e.g.,][]{BarnesSofia1996,Bouvier1997,Tinker2002}.  Longer lived disks allow for the star to lose more angular momentum through magnetic disk-locking \citep[e.g.,][]{Koenigl1991,Shu1994} or through accretion powered stellar winds \citep[e.g.,][]{Korycansky1995,Papaloizou1997,Matt2012}; stars with long-lived disks tend to be slower rotators than stars with short-lived disks \citep{Meibom2007}.  Binary companions at $\leq$ 100~AU shorten the disk dissipation timescale \citep{Lin1993,Artymowicz1994,Armitage1996}, causing angular momentum loss mechanisms such as accretion powered stellar winds and magnetic disk-locking will be halted, allowing for stars in close binary pairs to remain rotating faster for longer.  Close binary pairs with separations $<$10 AU will also undergo the effects of tidal forces, and over time will have their spin rate synchronized to the orbital motion \citep[e.g.,][]{Meibom2005},  increasing the stellar rotation.  Therefore, stars in binary systems with separations $<$100 AU should rotate faster due to tidal effects and/or disk truncation mechanisms.  If rotation indeed plays an important role in the production of magnetic acitivity of low-mass stars, then dMs in close binary pairs should be more active than their single star counterparts.

Some previous studies have found evidence for increased magnetic activity in close, eclipsing dM binaries \citep[][and references therein]{Morales2010}.  However, the paucity of low-mass eclipsing systems precludes any statistically significant study that focuses on how a close companion affects the observable properties of dMs.  Due to the intrinsically faint nature of dMs, large samples of both single or binary stars were historically untenable. However, with the advent of large, all-sky surveys (e.g., Sloan Digital Sky Survey \citep[SDSS,][]{Abazajian2009}, Two Micron All Sky
Survey \citep[2MASS,][]{Skrutskie2006}, UKIRT Infrared Deep Sky Survey (UKIDSS\footnote{The UKIDSS project is defined in \cite{Lawrence2007}. UKIDSS uses the UKIRT Wide Field Camera \citep[WFCAM][]{Casali2007} and a photometric system described in \cite{Hewett2006}. The  pipeline processing and science archive are described in Irwin et al
  (in prep) and \cite{Hambly2008}. We have used data from the 2nd data
  release, which is described in detail in \cite{UKIDSS}.}), NASA's Galaxy Evolution Explorer \citep[GALEX,][]{Martin2005}, etc.), large spectroscopic and photometric catalogs of low-mass stars have emerged.  SDSS alone has enabled catalogs that contain over 70,000 visually
confirmed dM spectra \citep[][hereafter W11]{W11} and more than 30 million 5-band images \citep{Bochanski2010}, providing statistically
robust samples of field dMs for Galactic and stellar studies.  While large samples of wide binary companions have been discovered in SDSS \citep{Dhital2010}, identifying large samples of close binary companions without high-resolution spectroscopy or adaptive optics imaging is challenging.  Several SDSS studies have identified samples of close dM pairs, but have not examined their effects on the measured magnetic activity properties \citep[e.g.,][]{Becker2011,Clark2012}.

One technique for identifying close companions to dMs is to focus on systems containing an dM and a white dwarf (WD), which can be easily identified in low-resolution spectroscopy.  Previous SDSS spectroscopic studies used
close and unresolved WD$+$dM binary pairs to investigate the magnetic
activity of dMs with close companions (e.g.,~\citealt{Raymond2003};~\citealt{S06}, hereafter S06;~\citealt{Heller2009}).  S06 compared the activity
fraction of 747 close WD$+$dM binary pairs as function of spectral type
to that of isolated dMs (W04).  In comparison, S06 found that the activity in early-type dMs is
higher in close binary pairs than in isolated dMs; similar to the result of \cite{Heller2009} who found 23.2\% of early-type unresolved white dwarf-main sequence (WDMS) to be active compared to $\sim$10\% of active early-type dMs from \cite{W04}.   From the WD cooling ages, S06 found that all of the active, early-type dM were found around young WDs.   There was some evidence that late-type dMs (in WD$+$dM pairs) also had a higher activity fractions, but there were not enough pairs to statistically confirm this result.  In addition, neither the \cite{Raymond2003} nor the S06 samples were large enough to put the WD$+$dM pairs in their proper Galactic context (how the magnetic activity varies with location in the Galaxy).  Additional SDSS data releases have dramatically increased the sample sizes of WD$+$dM pairs \citep{Rebassa2010,Rebassa2012} , which now allow for statistical investigations of the magnetic activity properties of low-mass stars in close binaries.

In this paper, we use a large sample of WD$+$dMs to build on the S06 analysis and investigate the magnetic activity properties of close WD$+$dM pairs as a function of spectral type, Galactic height (a proxy for age), WD cooling ages, and binary
separation.  In Section~2 we describe our sample selection and general properties of our sample. In Section~3 we describe the
procedures we developed to separate the binary components and calculate WD$+$dM
parameters.  We also present a detailed analysis of the WD spectra using cooling tracks to estimate age, mass, and temperature.  In addition, Section~3 details a radial velocity (RV)
analysis using individual SDSS exposures to uncover a population of very close binary pairs. In Section~4, we present our results on
activity fractions, lifetimes, and strength as a function of binary, stellar and Galactic parameters.  Section~5 discusses the results from our analysis and the potential implications of having close companions on the magnetic activity of dMs.  Finally, we summarize our findings in Section~6.

\section{Data}\label{Sec:Data}
We identified 1756, high-quality,  close WD$+$dM spectroscopic binaries from the SDSS Data Release 8 (DR8).  Our goal was not to produce a complete sample of WD$+$dM binary pairs (something that SDSS spectral selection precludes), but rather produce a sample of high
fidelity and high signal-to-noise (S/N) spectroscopic WD$+$dM binary pairs,
expanding upon previous close WD$+$dM catalogs: the
\cite{Rebassa2010,Rebassa2012} sample of 1602 and 2248 WDMS
pairs, respectively; and the S06 catalog of 747 close WD$+$dM
pairs.  In particular, we were interested in increasing the number of late-type dMs
(M7-M9) found by previous studies in an attempt to better understand how
activity in close pairs affects fully convective M dwarfs at the bottom of the main sequence.  We employed the use of multi-wavelength
color-cuts (ultraviolet, optical and infrared) to produce a sample of 1756 close WD$+$dM spectroscopic binaries
selected from the SDSS DR8 \citep[DR8,][]{Abazajian2009}
using the \cite{Rebassa2010}
WDMS sample as a guideline for developing our own color cuts
(described below).  The advantage of using a photometric selection criteria that spanned from ultraviolet (UV) to infrared (IR)
wavelengths was an increased sensitivity to late-type pairs
at infrared wavelengths, whose optical fluxes are dominated by the WD
components.  Inversely, this multi-wavelength technique was also sensitive to pairs with
very faint WD components whose optical flux is dominated by the dM.  In the end, our sample of
1756 close WD$+$dM pairs does not contain as many pairs as the \cite{Rebassa2012}
sample due to our stringent quality cuts that ensure high
S/N for both the WD and dM component; which were necessary for
our magnetic activity analysis.

We used photometry from four different public catalogs, spanning the
ultraviolet, optical, and infrared wavelengths, in addition to the optical
spectra from SDSS DR8 catalog.  The {\it fuv} (1350--1750~{\AA}) and {\it nuv}
(1750--2800~{\AA}) bands from GALEX; the $ugriz$ bands
(3000--11000~{\AA}) from SDSS \citep{Fukugita1996,Gunn1998}; and the {\it J} (1.25$\mu$m), {\it H}
(1.65$\mu$m), and {\it K} (2.15$\mu$m) band from both 2MASS and UKIDSS
were used. In addition to slightly different filters, 2MASS
\citep[Vega-based;][]{Skrutskie2006} and UKIDSS
\citep[AB;][]{Hewett2006} use different magnitude systems. While 2MASS
is an all-sky surey, UKIDSS probes a much smaller area but goes three
magnitudes deeper, creating an effective volume 12 times larger than 2MASS. Therefore, using both
2MASS and UKIDSS maximizes the number of WD$+$dMs that are detected in
the infrared.

\subsection{Sample Selection}
We selected all SDSS
``point sources'' with spectra that satisfied the following quality cuts: {\it ugriz}  $<$ 24 and redshift $<$
0.003.  Our initial leniency helped facilitate a large number of candidate pairs, while leaving room to filter out low quality objects in later
stages of the sample selection.  After removing
duplicate objects, our initial sample was composed of 520,701 SDSS  stars.  Multi-wavelength counterparts to these objects
were found in GALEX, 2MASS, and UKIDSS using three arcsecond matches to their positions. There were 227,728 matches in GALEX, 300,665 matches in 2MASS, and
131,416 matches in UKIDSS.

\begin{figure}[!ht]
\begin{center}
\includegraphics[trim=1cm 1cm 1cm 1cm,width=0.45\textwidth]{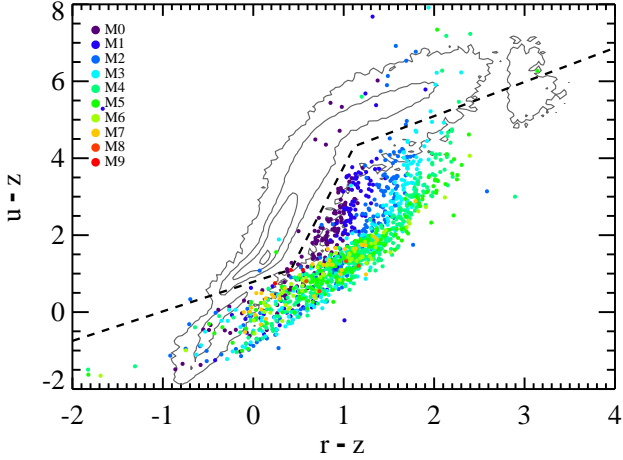}
\end{center}
\caption{{\it u -- z} versus {\it r -- z} color-color diagram for stars
  in SDSS.  The colored filled circles represent known WD$+$dM pairs from
  \cite{Rebassa2010}.  The symbols are colored according to the spectral
  type of the M dwarf.  The color cut is represented by the
  black-dashed line (all objects below that line are
  selected).  The grey contours are all of the 520,701 stars in the
  initial SDSS sample; the contour levels are 10, 100, 1000, 5000, and
  10000 objects. Since all of our objects have SDSS colors, these
  colors provide the most comprehensive cut.  We
  choose {\it r -- z} as its sensitive to dM spectral type while {\it
    u -- g} isolates the WD$+$dM population from the main stellar locus
  \citep{Smolcic04}.}
\label{fig:rz_uz}
\end{figure}

\begin{figure*}[!ht]
\begin{center}$
\begin{array}{cc}
\includegraphics[trim=1cm 0.5cm 0.5cm 1cm,width=0.5\textwidth]{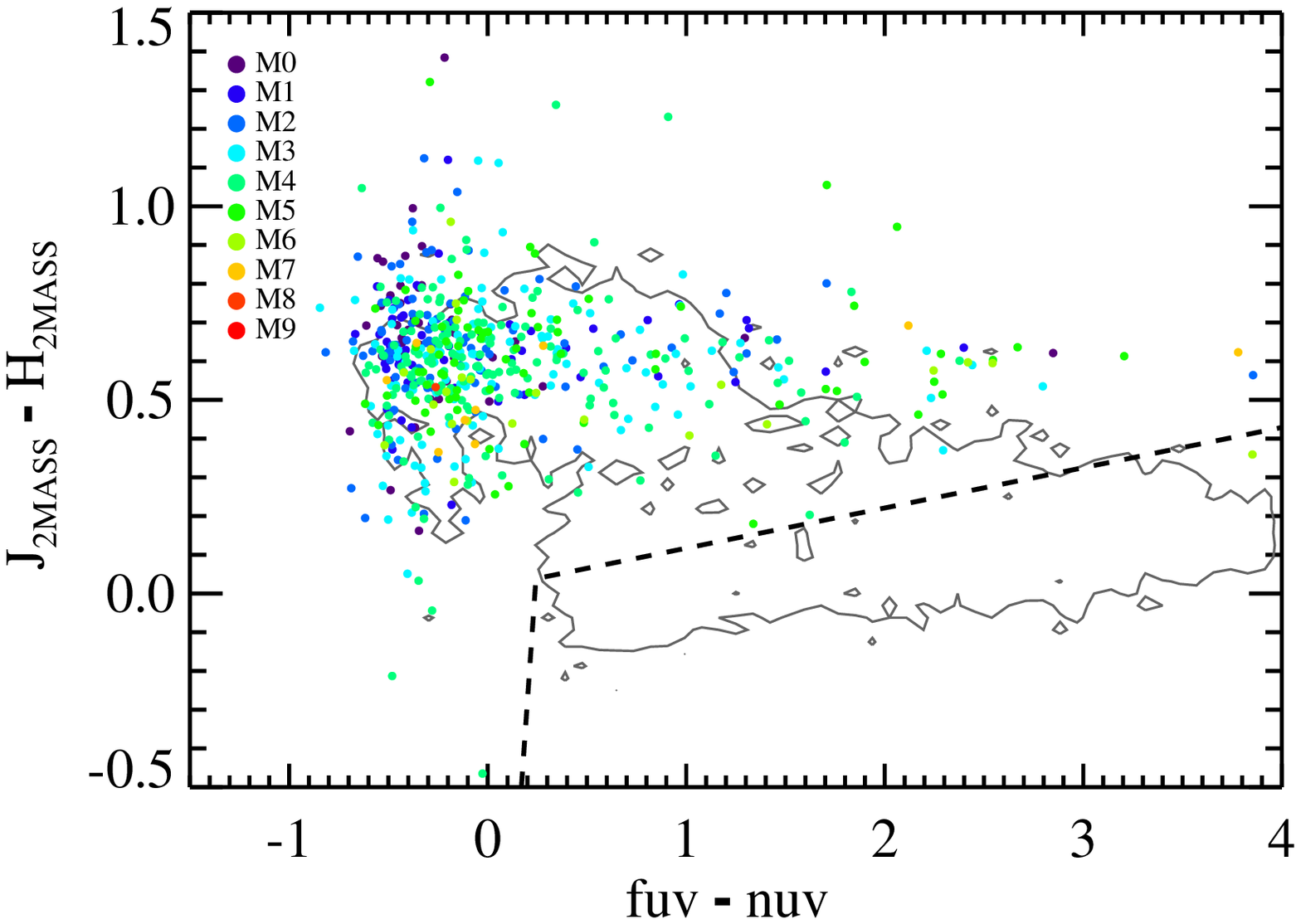} 
\includegraphics[trim=0.5cm 0.5cm 1cm 1cm,width=0.5\textwidth]{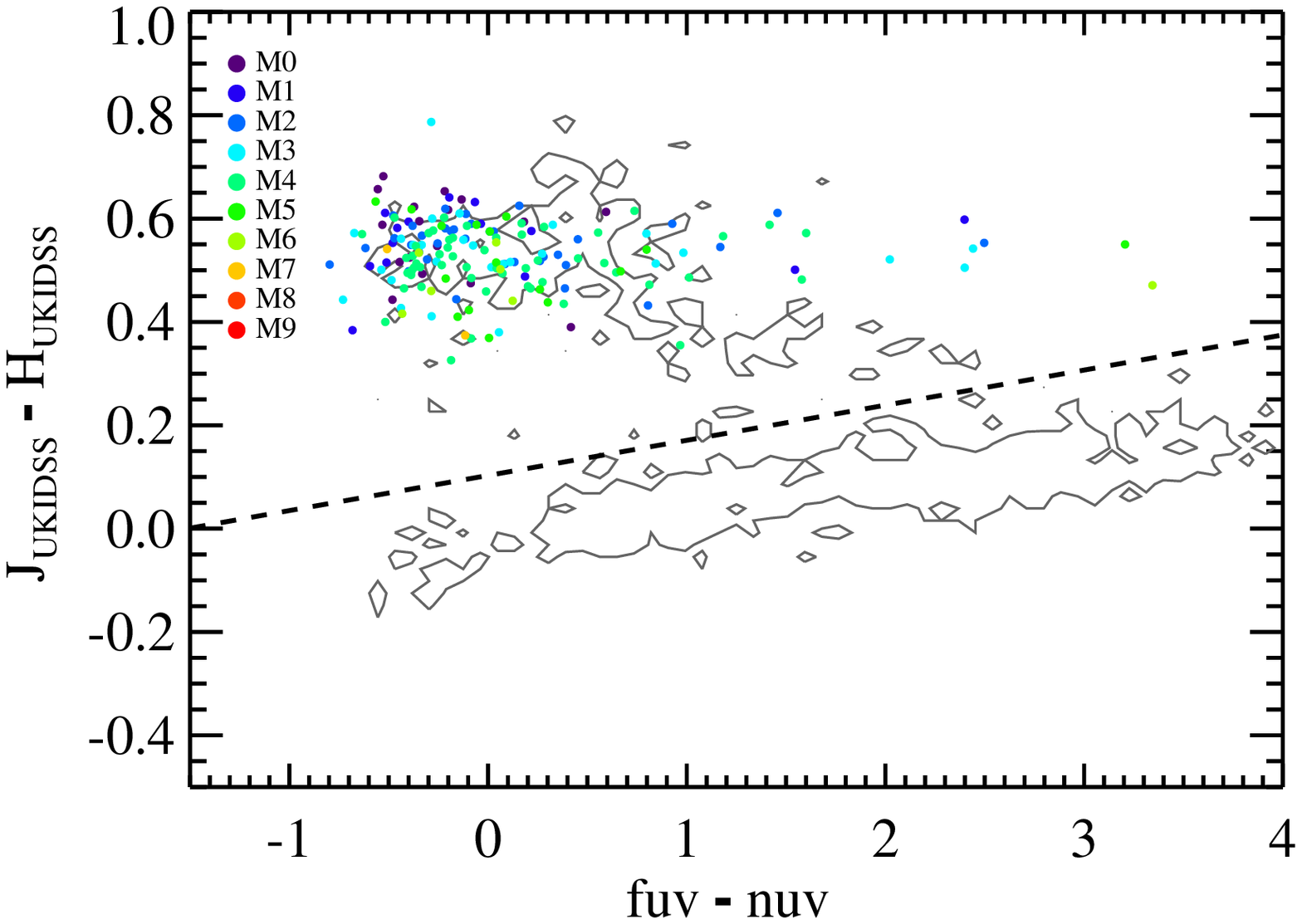} \\
\end{array}$
\end{center}
\caption{{\it fuv} -- {\it nuv} vs. {\it J} -- {\it H} color-color
  diagram.  As
in Fig. 1, the filled circles are colored by M dwarf spectral type for known WD$+$dM
pairs.  On the left panel, the contour represent the
2MASS data and there is only one contour levels representing stellar
densities $>$5.  The right panel uses UKIDSS data and only uses
contour levels with stellar densities $>$3.  The distinction was made because of the
photometric depth and precision disparity between the two samples.  This cut was
chosen so that the {\it fuv} -- {\it nuv} colors would isolate
new dM flux dominated pairs, where the WD may not component may not be
obvious in optical photometry but will show up in the UV regime.
Similarly, {\it J} -- {\it H} selects WD flux dominated pairs where
the dM will reveal itself in the infrared.  {\it J} -- {\it H} is also sensitive to late-dM spectral types.}
\label{fig:fuvnuv_jh}
\end{figure*}

We reduced the number of false positive candidate WD$+$dM binary systems
by enforcing additional color-cuts in {\it fuv}, {\it nuv}, {\it u},
{\it g}, {\it r}, {\it i}, {\it z}, {\it J}, {\it H}, and {\it K}
color space.  We used the color cuts provided by \cite{Rebassa2010} as
an initial guide for our selection.  We visually inspected the 1602
WDMS spectra from \cite{Rebassa2010} and concluded that 1399 were
WD$+$dM pairs.  We then explored the color space defined by all of the
above filters and traced out regions with minimimal contamination
(we defined contamination to mean the percentage of non WD$+$dM pairs
included in a color cut).  Our final color cuts were similar to those
described in the \cite{Rebassa2010} analysis, but were more useful in probing later spectral types (M7-M9) as well as
spectra dominated by WD or dM flux. Our color cuts were applied to
the latest SDSS DR8 catalog, where we found 364 new pairs that were
not present in \cite{Rebassa2010}.  For the following discussion we
will compare the effectiveness of our color cuts to the
\cite{Rebassa2010} sample (on which we base our sample selection).  Since we began our analysis, a larger
sample of WD$+$dM pairs has been published \citep{Rebassa2012}.  We will
compare our sample to that of \cite{Rebassa2012} in a later section.
The color-cuts are shown in Figures~\ref{fig:rz_uz}-\ref{fig:nuvk_zj}.
In each figure, the contoured data represent all SDSS point sources,
while the colored {\it filled circles} represent confirmed WD$+$dM from
\citep{Rebassa2010}.  The circles are colored according to their
spectral type; this was done in an effort to identify trends in the
spectral types in order to best select regions hosting the later M7-M9
spectral types that have not been well sampled in previous studies.

When using  {\it J}, {\it H}, and {\it K} colors, we
needed to distinguish between 2MASS \citep{Hewett2006} and UKIDSS
\citep{Skrutskie2006} photometry separately due to the differences in the photometric
precision, the zeropoints, and the filter responses.  Treating them
separately did not identify any significant trends that could be due to using slightly different photometric systems. However, we will present our 2MASS
and UKIDSS color cut analysis separately below.

\subsection{Color Cuts}
WD$+$dM systems are well separated from main sequence stars in a {\rz}
vs. {\uz} color--color diagram \citep{Smolcic04}. This is largely due to the fact that the {\uz} color spans the wavelengths where the WD and dM components peak.  The {\rz} color serves to
segregate the M dwarfs as a function of spectral type. 
Figure~\ref{fig:rz_uz} shows the {\rz} vs. {\uz} color--color diagram
for all SDSS stars. The main-sequence stars are shown as grey
contours, while the WD$+$dM systems spectroscopically identified in \cite{Rebassa2010}
are shown as circles, with the colors corresponding to spectral type
of the dM. The WD$+$dM population is clearly segregated from the
main-sequence stars in {\uz}. We did a by-eye
fit to separate the WD$+$dM pairs from the main stellar locus, shown as
dashed lines in Figure~\ref{fig:rz_uz} and described by:
\begin{eqnarray}\label{Eq:rzuz}
(u-z) < \left\{ \begin{array}{ll}
    0.767&(r-z) + 0.786; \\ &\textrm{for} ~(r-z) <0.416, \\ 
    4.573&(r-z) - 0.798; \\ &\textrm{for}~0.416 < (r-z) < 1.117, \\ 
    0.888&(r-z) + 3.318; \\ &\textrm{for}~(r-z) > 1.117. 
    \end{array}\right\}
\end{eqnarray}
As can
be seen in Figure~\ref{fig:rz_uz}, this selects the vast majority of
WD$+$dM pairs. However, systems that are dominated by either the WD
or dM component might not be separated in the optical
wavelengths. Using the UV (where the WD dominates) and the
near-IR (where the dM dominates) separates those WD$+$dM systems
more cleanly. Therefore, we used the UV data from GALEX and the
near-IR 2MASS and UKIDSS to identify additional pairs.

\begin{figure*}[!ht]
\begin{center}$
\begin{array}{cc}
\includegraphics[trim=1cm 0.5cm 0.5cm 1cm,width=0.5\textwidth]{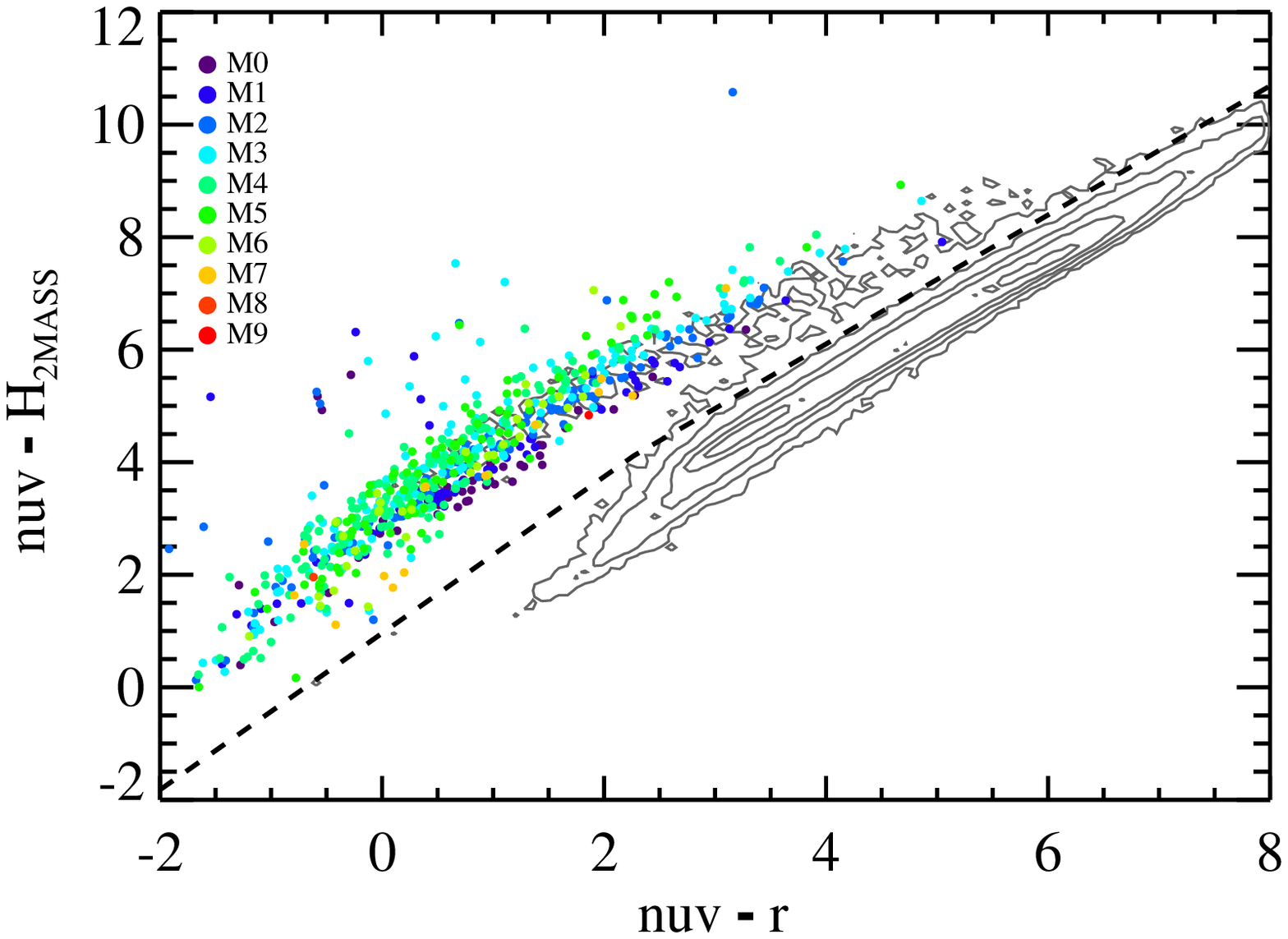} 
\includegraphics[trim=0.5cm 0.5cm 1cm 1cm,width=0.5\textwidth]{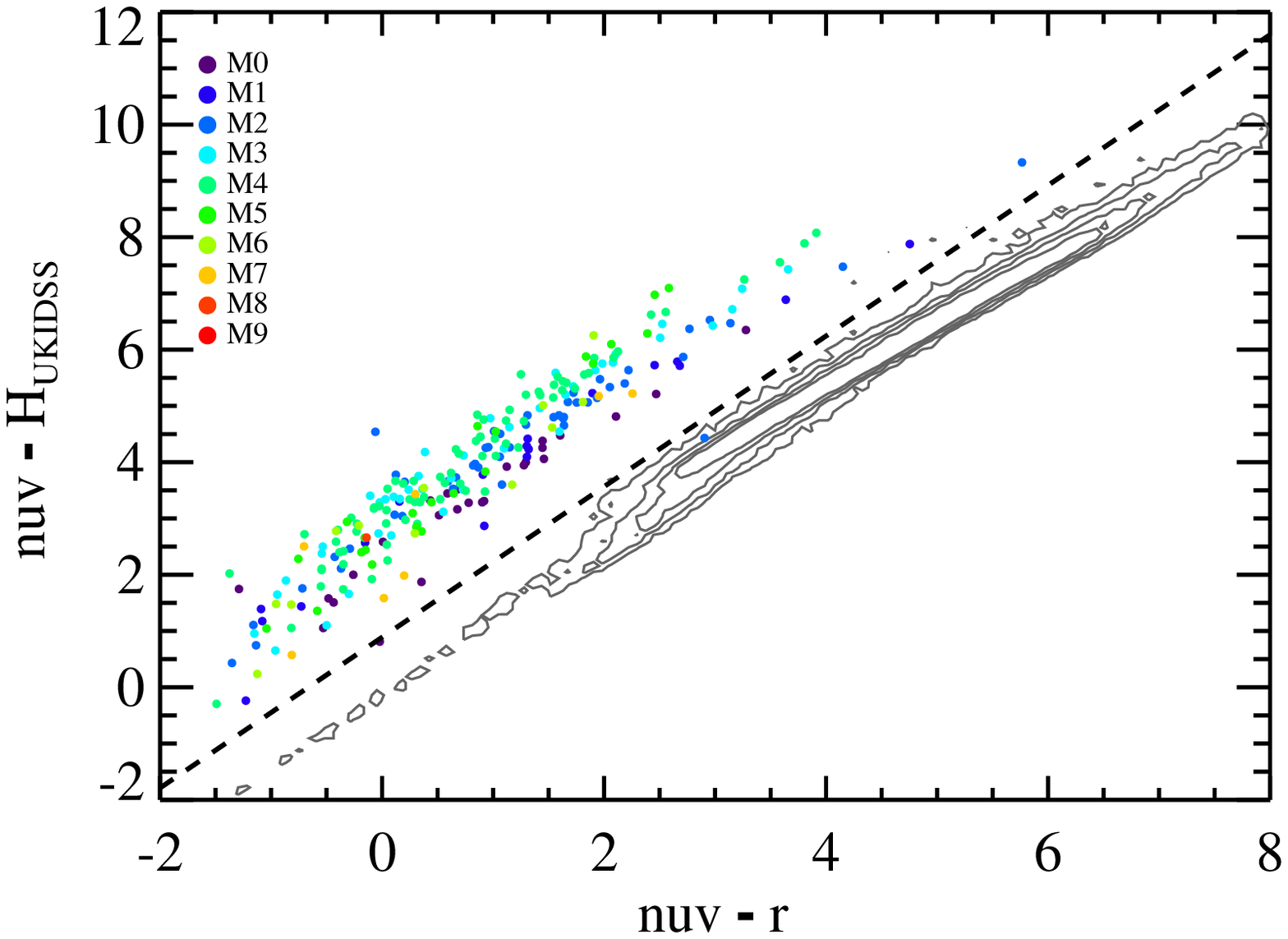} \\ 
\end{array}$
\end{center}
\caption{{\it nuv} -- {\it r} vs. {\it nuv -- H} color-color diagram,
  filled circles are colored by spectral type as before as well as distinguishing
between 2MASS and UKIDSS separately.  The contours levels on the right
panel represent stellar densities of 5, 25, 100, 250, and 500 in the
2MASS data and 5, 20, 50, and 100 in the UKIDSS data on the left
panel. This cut was chosen for its
ability to separate the WD$+$dM bridge \citep{Smolcic04} from
the main stellar locus.  Colors {\it nuv -- r} provides excellent spread in
spectral type, while {\it nuv -- H} helps select dMs with later spectral types.}
\label{fig:nuvr_nuvh}
\end{figure*}

Figure~\ref{fig:fuvnuv_jh} shows the {\fnuv} vs. {\JH} plot for our sample.  The {\fnuv} isolated objects with significant UV flux, while {\JH} color separated
the M dwarfs from the G and K dwarfs. While this cut
is similar to the {\fnuv} vs. {\rz} cut used in \cite{Rebassa2010},
using the {\JH} color, instead of {\rz} color, produced a cleaner
separation. The locus of the WD$+$dM pairs was described by 

\begin{eqnarray}\label{Eq:JH1}
(J - H)_{\rm 2MASS} > \left\{ \begin{array}{ll}
    7.378&({\fnuv}) - 1.754; \\ &\textrm{for} ~({\fnuv}) < 0.243, \\ 
    0.104&({\fnuv}) + 0.013; \\ &\textrm{for}~({\fnuv}) \ge 0.243, \\ 
    \end{array}\right\}
\end{eqnarray}
for 2MASS colors and by 
\begin{eqnarray}\label{Eq:JH2}
(J - H)_{\rm UKIDSS} > 0.068~({\fnuv}) + 0.103
\end{eqnarray}

\begin{figure*}[!ht]
\begin{center}$
\begin{array}{cc}
\includegraphics[trim=1cm 0.5cm 0.5cm 1cm,width=0.5\textwidth]{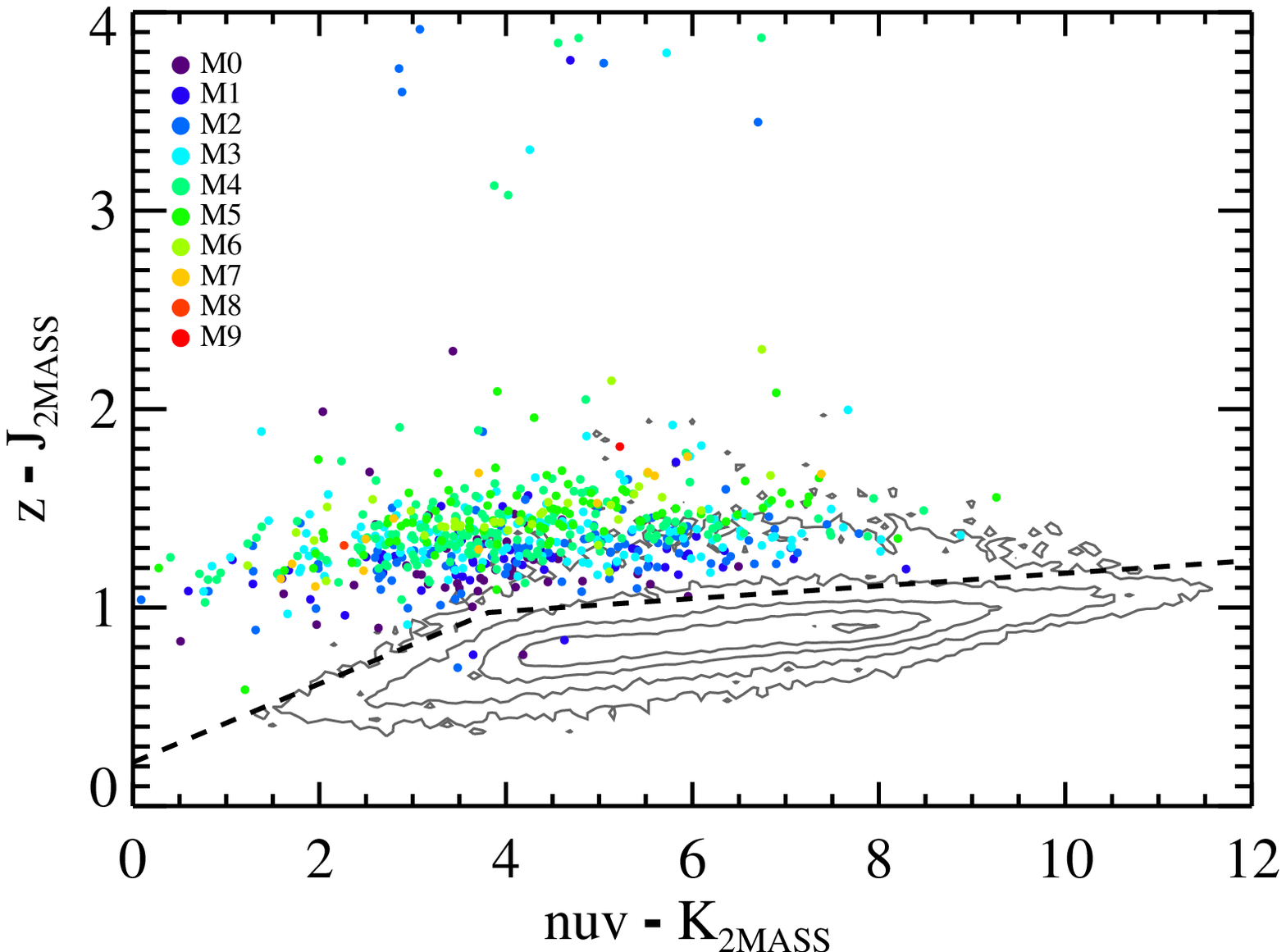} 
\includegraphics[trim=0.5cm 0.5cm 1cm 1cm,width=0.5\textwidth]{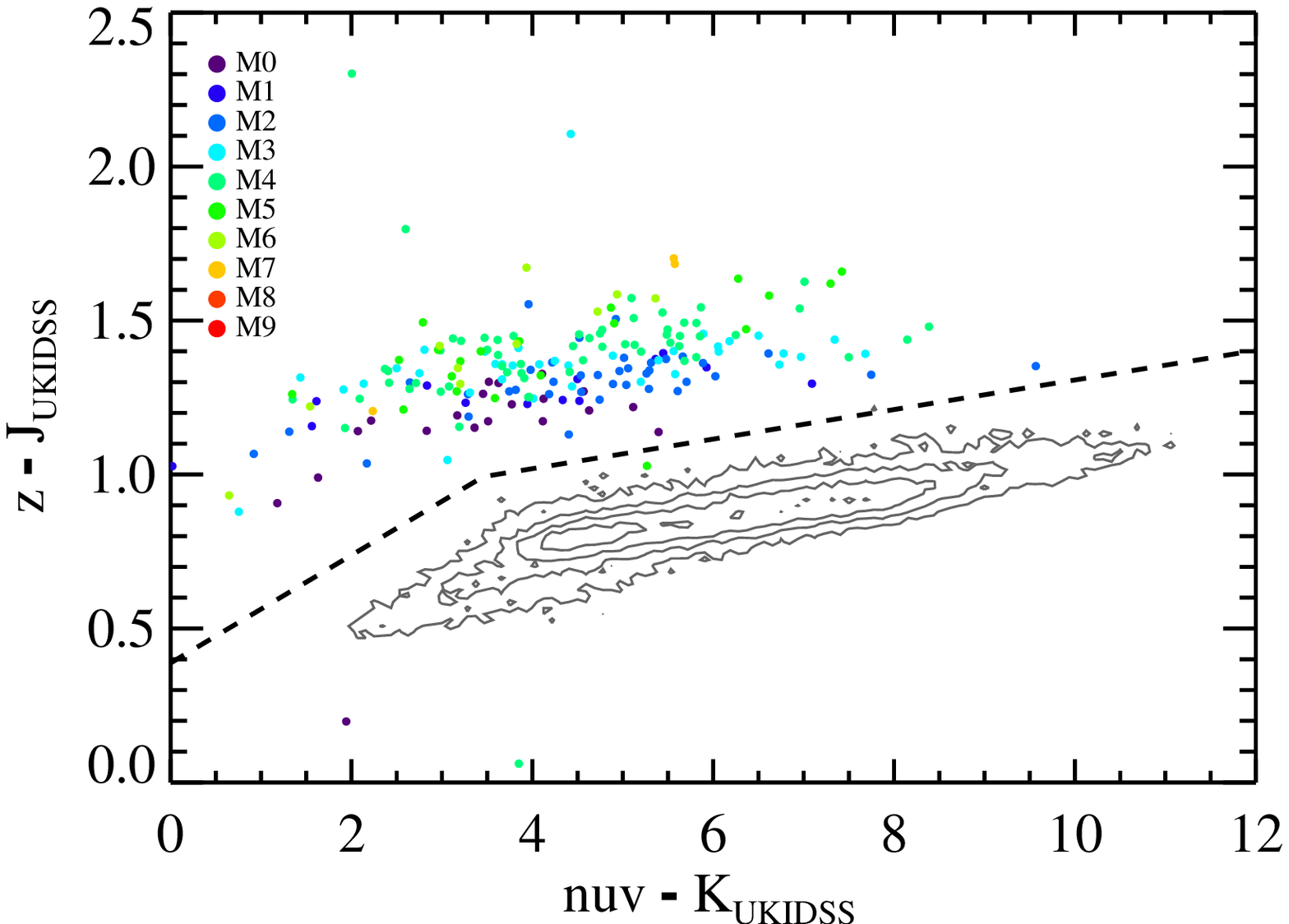} \\ 
\end{array}$
\end{center}
\caption{{\it nuv} -- {\it K} vs. {\it z} -- {\it J} color-color diagram.
   Filled circles are colored by M dwarf spectral type.  Left and
  right panels distinguish between 2MASS and UKIDSS data
  respectively. The contours levels on the right
panel represent stellar densities of 5, 25, 100, 250, and 500 in the
2MASS data and 5, 20, 50, and 100 in the UKIDSS data on the left
panel.
  These cuts were chosen specifically
to identify late spectral types with {\it z} -- {\it J}, which
covers the wavelength regime where there is a peak in M7-M9 flux.}
\label{fig:nuvk_zj}
\end{figure*}
for UKIDSS colors. We then used the {\it nuv} $-$ {\it r} vs. {\it nuv} $-$ {\it H}
diagram to further separate WD$+$dM pairs from the stellar locus. As 
Figure~\ref{fig:nuvr_nuvh} shows, WD$+$dM pairs are above the
main-sequence stars in {\it nuv} $-$ {\it H} while {\it nuv} $-$ {\it r} 
spreads the pairs as a function of the dM spectral type. This cut is
only slightly different from the  {\it nuv} $-$ {\it i} vs. {\it nuv}
$-$ {\it H} used in \cite{Rebassa2010} and is a better tracer of dM
spectral type. The derived WD$+$dM locus is described by
\begin{eqnarray}\label{Eq:NUV1}
nuv - H_{\rm 2MASS} > \left\{ \begin{array}{ll}
   1.391&(nuv - r)+0.960; \\ &\textrm{for}~(nuv-r) <2.226, \\ 
   1.147&(nuv - r)+1.513; \\ &\textrm{for}~(nuv-r) \ge 2.226, \\ 
    \end{array}\right\}
\end{eqnarray}
for 2MASS colors and 
\begin{eqnarray}\label{Eq:NUV2}
nuv-H_{\rm UKIDSS} > 1.339 \times (nuv - r) + 0.888.
\end{eqnarray}
for UKIDSS colors. As the fourth, and final, color cut, we used 
{\it nuv} $-$ {\it K} vs. {\it z} $-$ {\it J}, which is only slightly
different from the third one we used. However, {\it z} $-$ {\it J} is
more sensitive to systems with later-type dM secondaries, as it
includes the wavelength regime where late-dMs
peak in flux.  Figure~\ref{fig:nuvk_zj} shows the WD$+$dM locus well separated above the main stellar sequence. The derived WD$+$dM locus
is described by
\begin{eqnarray}\label{Eq:zJ1}
z - J_{\rm 2MASS} > \left\{ \begin{array}{ll}
    0.198~(nuv - K_{\rm 2MASS})+0.220; \\ \quad~~ \textrm{for}~(nuv - K_{2MASS}) < 3.813, \\ 
   0.032~(nuv - K_{\rm 2MASS})+0.853; \\ \quad~~ \textrm{for}~(nuv - K_{2MASS}) \ge 3.813, \\ 
    \end{array}\right\}
\end{eqnarray}
for 2MASS colors and by 
\begin{eqnarray}\label{Eq:zJ2}
z - J_{\rm UKIDSS}  > \left\{ \begin{array}{ll}
    0.175~(nuv - K_{\rm UKIDSS})+0.388; \\ \quad~ \textrm{for}~(nuv - K_{\rm UKIDSS})< 3.457, \\ 
    0.048~(nuv - K_{\rm UKIDSS})+0.827; \\ \quad~ \textrm{for}~(nuv - K_{\rm UKIDSS}) \ge 3.457, \\ 
    \end{array}\right\}
\end{eqnarray}
for UKIDSS colors. 

Using the color cuts above, we selected
53,279 unique objects.  The color cuts were able to recover 1379 of
the 1399 WD$+$dM pairs from \cite{Rebassa2010}; the pairs that were missed by our color cuts had inconsistent colors with the WD$+$dM locus (seen in Figure~\ref{fig:rz_uz}).  Each object was
visually inspected and given 
one of three simple classifications: not a WD$+$dM, a high probability
candidate, or a low probability candidate.  The first two classifications
are self explanatory and the third was necessary for objects that
were difficult to differentiate from single WDs or dMs with low
S/N or other astronomical objects.  After classifying each object by-eye, there were 1,891 high-probability WD$+$dM pairs and 2,220
low-probability pairs.  We performed a detailed spectral analysis
(described in Section~3) on all of the candidate pairs and were able
to confirm that of the 1,891 high-probability pairs 1,460 were real
and of the 2,220 low-probability pairs 199 were real, resulting in an
initial sample size of 1659 WD$+$dM pairs.  The {\rz} vs. {\uz} cut selected
1605 of the WD$+$dM while the IR and UV color cuts were able to select
the remaining 54 pairs that were not included in the {\rz} vs. {\uz} cut
(Equation~\ref{Eq:rzuz}).  Of those 54 remaining pairs, only one pair was found
exclusively using the {\fnuv} vs. {\JH} cuts (Equations~\ref{Eq:JH1}
and \ref{Eq:JH2}), 18 pairs were found exclusively from the 
{\it nuv} $-$ {\it r} vs. {\it  nuv} $-$ {\it H} cuts (Equations~\ref{Eq:NUV1} and
\ref{Eq:NUV2}), and three pairs were found exclusively from the {\it nuv} - {\it
  K} vs. {\it z} -{\it J} cuts (Equations~\ref{Eq:zJ1} and \ref{Eq:zJ2}).

In creating the \citetalias{W11} M dwarf sample, over 70,000 dM
were visually confirmed from over 116,000 candidate spectra from SDSS
DR7.  During the visual inspection process, candidate WD$+$dM objects
were flagged for later analysis, from these we added 194 possible
WD$+$dM that were not in our sample.  In addition, we matched our sample
to those of \cite{Rebassa2010} and S06 and found 522 and 144
objects, respectively, that we had missed.  This resulted in a
total of 910 candidate WD$+$dM objects that we had missed in our color cut
analysis.  After visually inspecting each spectrum, we added 141
objects to our sample. The other 769 objects were eliminated as they were
non-WD$+$dM pairs, WD or dM stars with low S/N in the red or
blue portion of the spectrum (making separation of a companion
difficult), or WD$+$dM pairs with S/N too low for our magnetic activity
analysis.   Finally, we did a final inspection of our sample
and found 44 objects that had been misclassified or where the S/N was less
than 3 near H$\alpha$ absorption line and therefore did not meet 
our quality criteria.  This resulted in a final sample of 1,756 confirmed close
WD$+$dM pairs.  As we have mentioned earlier, this sample was not
devised to be complete but rather a sample of high fidelity and high
signal-to-noise spectroscopic WD$+$dM pairs.  The complete catalog will
be available online.\footnote{http://vizier.u-strasbg.fr/}\footnote{http://delia.phy.vanderbilt.edu/slowpokes?d=wdm}

When we compared our final sample to the
S06 and \cite{Rebassa2010,Rebassa2012} catalogs, we found that we were
missing 310, 409, and 606 pairs, respectively. However, we added 
847, 556, and 121 new pairs, respectively, to those three samples.
The pairs missing from our sample were visually inspected; and the majority of
them were not WD$+$dM pairs but rather WD$+$K pairs, had low S/N, or
the components could not be properly separated with our spectral
analysis procedure (described in Section~3.1).  Thus, they were not
useful for our magnetic activity study and not included in the
final sample.  

\begin{figure}[!ht]
\begin{center}
\includegraphics[trim=1cm 1cm 1cm -0.25cm, width=.45\textwidth]{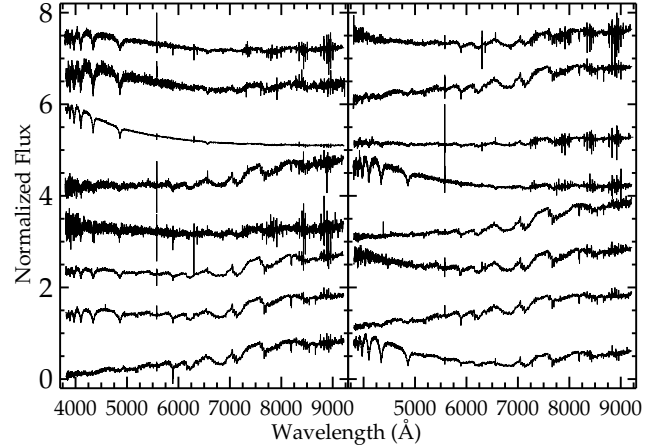}
\end{center}
\caption{Examples of a typical SDSS WD$+$dM spectra reported in
  this paper.  The fluxes have been normalized for
  visualization purposes.  Note that large feature commonly found
  at around 5500\AA\ is an artifact caused by the SDSS spectrograph
  and is not a real feature (S06).}
\label{fig:exspectra}
\end{figure}

\section{Analysis}
\subsection{WD$+$dM separation procedure}
\begin{figure*}[!ht]
\begin{center}$
\begin{array}{cc}
\includegraphics[trim=1cm 0.5cm 0.5cm 1cm, width=0.5\textwidth]{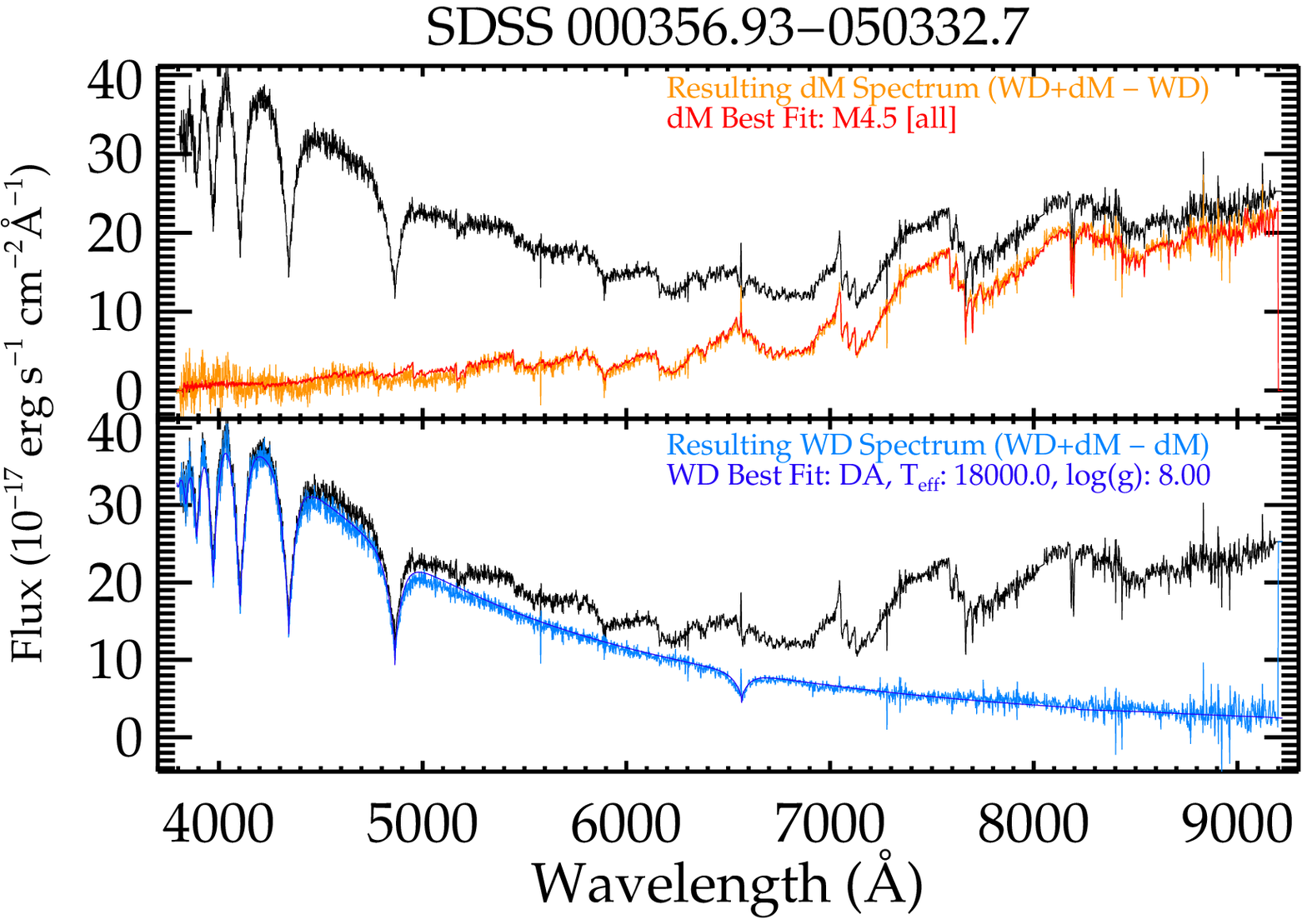} 
\includegraphics[trim=0.5cm 0.5cm 1cm 1cm, width=0.5\textwidth]{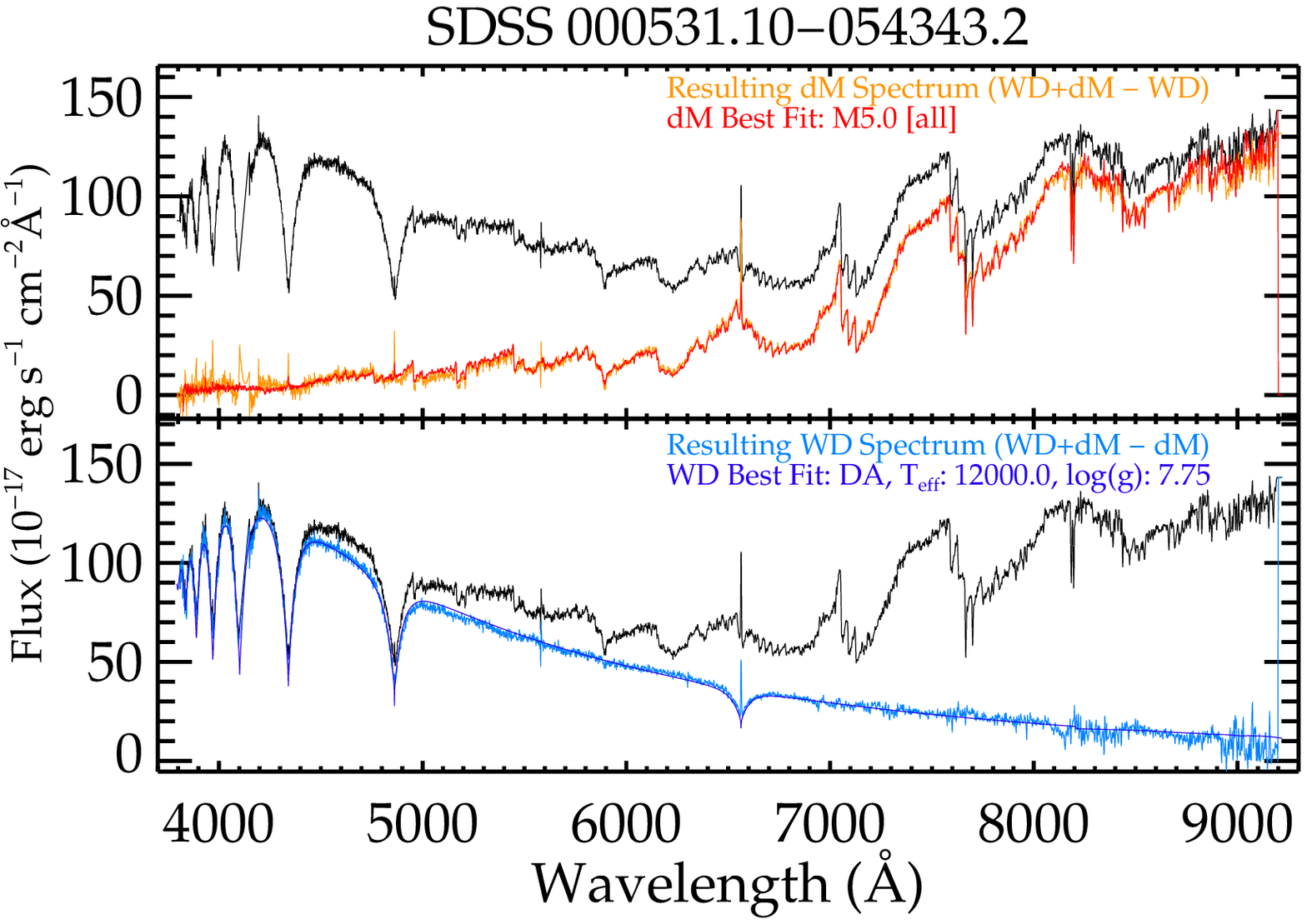} \\
\includegraphics[trim=1cm 1cm 0.5cm 0.5cm, width=0.5\textwidth]{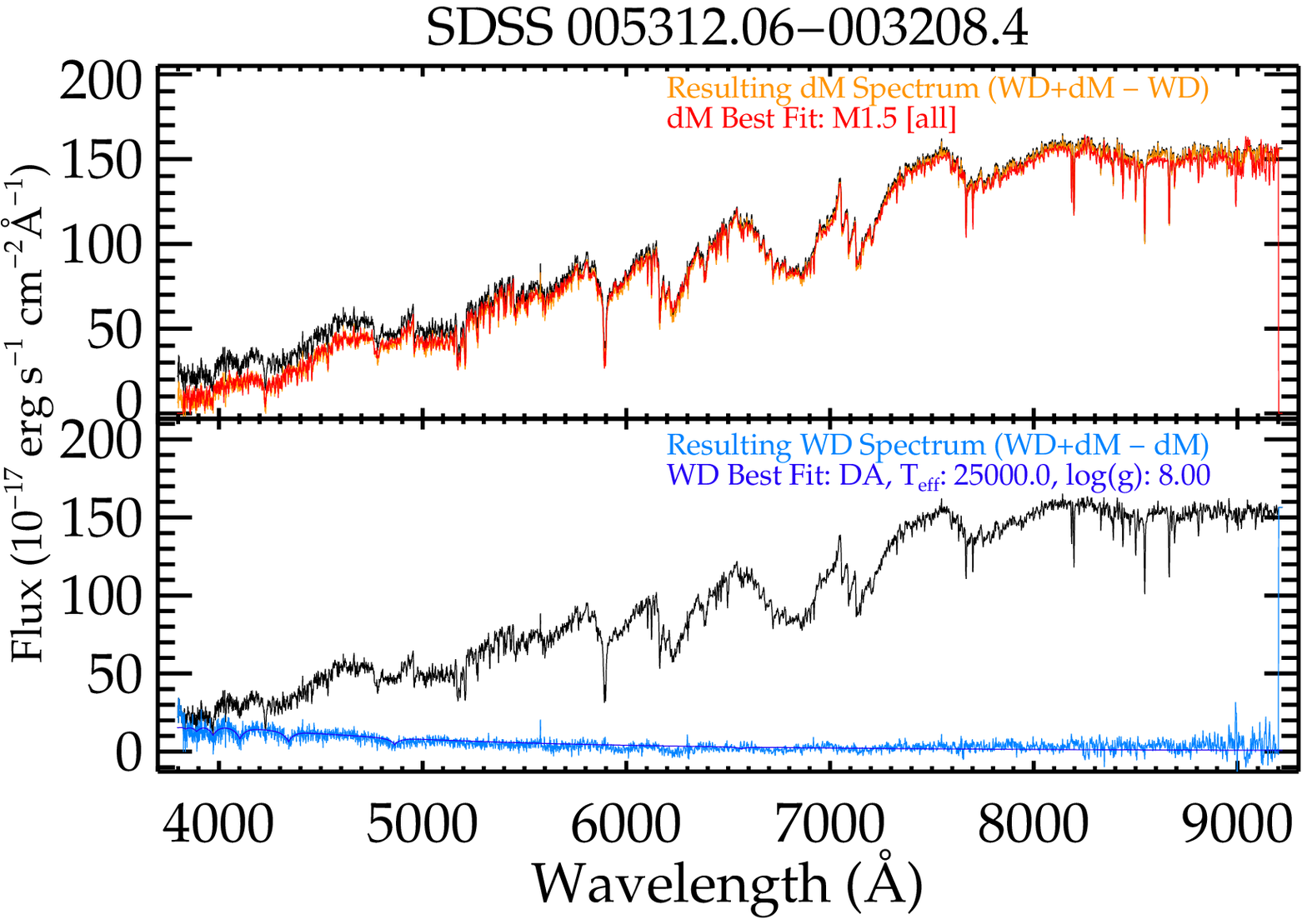} 
\includegraphics[trim=0.5cm 1cm 1cm 0.5cm, width=0.5\textwidth]{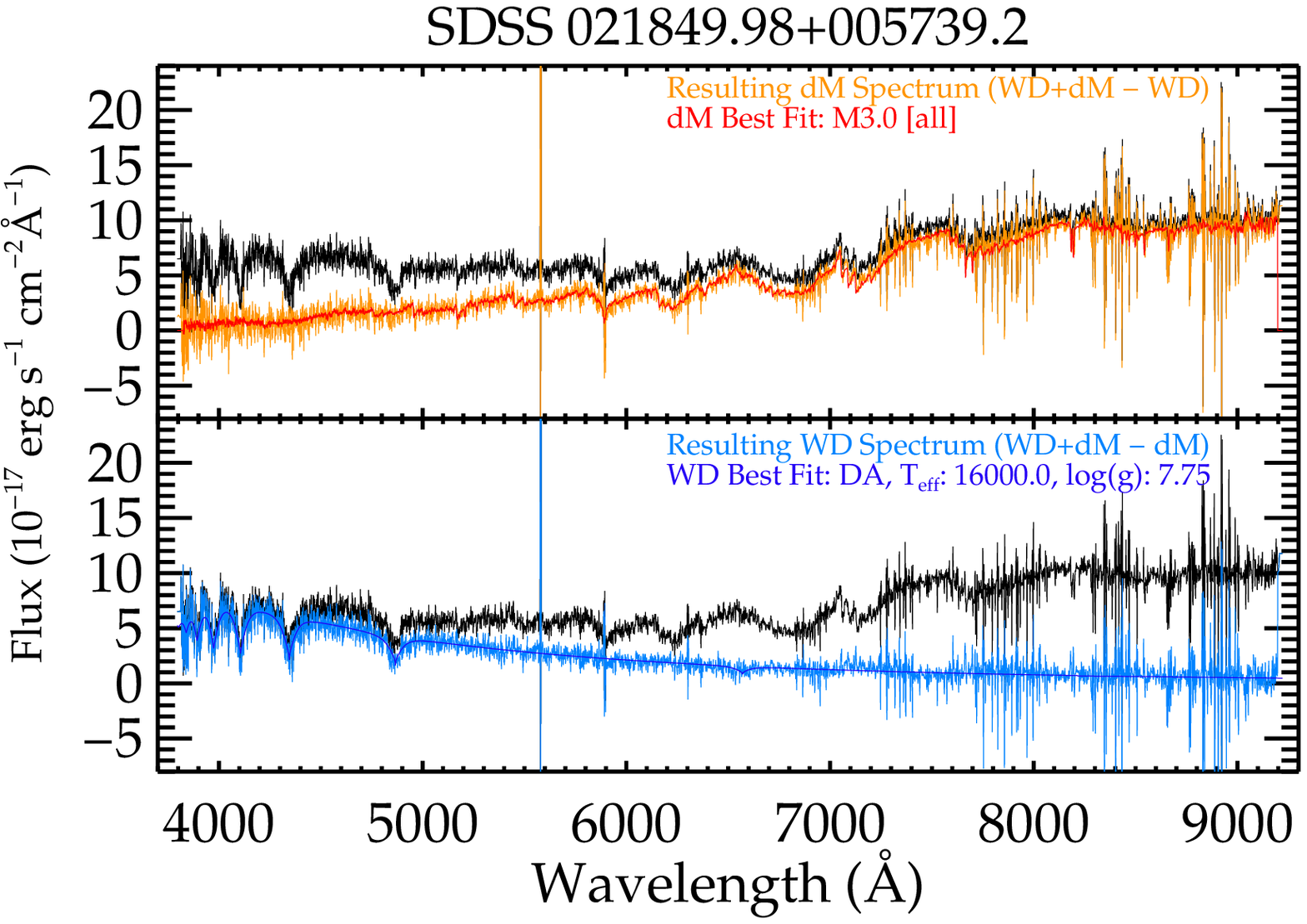} \\
\end{array}$
\end{center}
\caption{Four examples of the results from our WD$+$dM spectral separation
  procedure as outlined in Section~3.1.  Top panels show the original SDSS spectrum (black), with the
  recovered dM spectrum (orange; original SDSS spectrum
  subtracted by the best fit WD template), and the best fit dM
  template (red).  On the bottom panel is again the original SDSS spectrum (black), with the
  recovered WD spectra (light blue; original SDSS spectrum
  subtracted by best fit dM template), and the best fit WD model (dark
  blue).}
\label{fig:wddm_spectra}
\end{figure*}
Most of our sample consists of unresolved spectroscopic WD$+$dM
binaries (see in Figure~\ref{fig:exspectra} for 16 representative spectra from our sample).
To conduct a proper analysis of the individual stellar components, it is necessary to separate the
individual spectra from the combined spectrum.  To do this, we constructed an iterative method for separating the
combined WD$+$dM spectrum into its individual components by 
fitting the appropriate WD models \citep{Koester2001} and dM templates
\citep{Bochanski2007} to each spectrum.  The \cite{Koester2001}
WD models consist of 372 DA and DB WD models for varying temperatures
and $\log g$.  The DA models vary from 6,000
to 100,000 K and each temperature bin has a separate model for $\log g$
values of 5, 5.5, 6,
6.5, 7, 7.5, 8, 8.25, 8.5, and 9.  The DB white dwarf
models vary from 10,000 to 40,000K and have $\log g$ values of 7, 7.5, 8,
8.5, and 9.  We used half spectral type dM templates (M0, M0.5, M1,
M1.5\ldots M8.5, M9; J. Bochanski, {\em private communication}) that were derived from composites of high S/N spectra \citep{Bochanski2007}.

The iterative process was comprised of five steps: 1) We began by
fitting the combined spectrum (see Figure 5 for examples)
with the WD models and subtracting the best-fit $\chi^{2}$ model from the
combined spectrum;  2) we then fit a dM template to
the WD subtracted spectrum to find the best-fit dM template; 3) then,
we subtracted the best-fit dM template (from step 2) from the original combined
spectrum and then found the best-fit WD to the dM subtracted spectrum;
4) again, we found the best-fit dM template to the WD subtracted
spectrum; and 5) we repeated this process until the best-fit WD and dM converged (did not change after repeated iterations).  The process typically finished after only two or
three iterations, however some fits were degenerate and alternated
between different values of a single parameter.  For the few cases where solutions did not converge, we stopped the procedure after ten iterations and chose the model fit with the lowest
overall $\chi^{2}$ value.  We note that we do not distinguish between cold and hot WD solutions in our fitting process as it is not imperative for our magnetic activity analysis; this could be the cause of some of the solutions failing to converge in the fitting process.

For the dM fit, we used a best-fit
$\chi^{2}$ technique over a wavelength range of 7000-9300\AA.  
We found that evenly weighting the entire wavelength regime of the dM provided the most accurate fits.  However, the WD fitting
process was more complicated; we placed higher weights on the
hydrogen Balmer absorption lines and weighted them $\sim$1000 times higher than the continuum to ensure that both the shape and the
depth of the absorption lines were accurately fit, features important for determining $\log g$.  We focused on
H$\alpha$, H$\beta$, H$\delta$, and H$\gamma$ and weighted each
predicted peak (corrected for RV of the WD, detailed below) in addition to 100\AA\ region on 
either side of each peak to ensure the Balmer regions were fit well.  Additionally, we 
weighted 10\AA\ wide portions between the absorption lines $\sim$1000 times higher to assist in
properly fitting the overall
shape of the WD continuum.  This procedure resulted in 
accurate fits for both the Balmer lines as well as the
continuum, which ensured an accurate subtraction of the WD from the
combined SDSS spectrum and a clean dM spectrum.  We also took into account possible
hydrogen Balmer series emission from an active dM
component that could affect the WD fit.  We down-weighted the predicted locations of
dM emission (corrected for RV) $\sim$1000 times lower to minimize contamination.

\begin{deluxetable*}{lrlrllrlrr}
\tablewidth{0pt}
\tablecolumns{10}
\tabletypesize{\scriptsize}
\tablecaption{WD+dM Parameters: Spectral classifications and radial velocities.}
\tablehead{\colhead{SDSS ID} &
	   \colhead{Plate} &
	   \colhead{MJD} &
	   \colhead{Fiber} &
	   \colhead{dM}	&
	   \colhead{WD}	&
	   \colhead{WD T$_{eff}$} &
	   \colhead{WD} &
	   \colhead{dM RV} &
	   \colhead{WD RV} \\
	   \colhead{} &
	   \colhead{} &
	   \colhead{} &
	   \colhead{} &
	   \colhead{SpT} &
	   \colhead{Type} &
	   \colhead{(K)\tablenotemark{1}} &
	   \colhead{$\log g$\tablenotemark{1}} &
	   \colhead{(km s$^{-1}$)} &
	   \colhead{(km s$^{-1}$)} }
\startdata
SDSS J000109.42$+$255459.5  & 2822 & 54389 & 334 & M2.5 & DA & 16000 & 7.00 & -22.316 & 208.306 \\
SDSS J000152.10$+$000644.7  & 387   & 51791 & 157 & M0     & DA & 40000 & 7.50 & 4.076 & \nodata \\
SDSS J000250.65$-$045041.6 & 2630 & 54327 & 439 & M4    & DB & 20000 & 8.50 & -11.468 & -36.618 \\
SDSS J000356.93$-$050332.7 & 2630 & 54327 & 173 & M4.5 & DA & 18000 & 8.00 & -52.853 & 73.650 \\
SDSS J000421.61$+$004341.5  & 685    & 52203 & 561 & M5    & DB & 16000 & 9.00 & -22.385 & \nodata \\
SDSS J000447.78$+$291140.9  & 2824 & 54452 & 600  & M0.5 & DA & 16000 & 8.00 & -18.033 & -35.996 \\
SDSS J000453.94$+$265420.4  & 2824 & 54452 & 78    & M4    & DA & 10000 & 8.25 & -29.225 & 174.521 \\
SDSS J000504.92$+$243409.7  & 2822 & 54389 & 180  & M4.5 & DA & 11000 & 8.25 & 41.247 & 29.640 \\
SDSS J000531.10$-$054343.2  & 2624 & 54380 & 82 & M5 & DA & 12000 & 7.75 & -23.974 & -46.497 \\
SDSS J000559.87$-$054416.1  & 2624 & 54380 & 60 & M1.5 & DA & 30000 & 8.00 & -9.603 & 101.562 
\enddata
\tablenotetext{1}{We note that the values for WD $T_{\rm eff}$ and WD $\log g$ values are only estimates based on the best-fit models.  The WD parameters reported in~\ref{tab:wddm2} are calculated using methods described in Section~3.2 and are more reliable.}
\label{tab:wddm1}
\end{deluxetable*}

After each iteration, we used a cross-correlation technique to calculate the
RVs of the WD and dM components. The cross-correlation technique compared the separated WD and dM to their corresponding rest wavelength best-fit template; the shift in the cross-correlation peak yielded the component RV.  The measured RV was then used to shift
the wavelength axis of the template and provide a better fit in subsequent iterations.
For the cross-correlations, we used a wavelength range of 3900-5200\AA\ 
for the WD  (this regime was chosen to include H$\beta$, H$\gamma$, and
H$\delta$) and 6600-9000\AA~for the dM (this range
includes the strong Na \textsci $\lambda \lambda$ [8183.26, 8194.79\AA] doublet
and a number of TiO molecular
features).  In an effort to determine the uncertainty
in the measured RV, we chose two bins over the
wavelength ranges designated above for the WD and four bins for the dM
component.  For example, the WD had two bins 3900-4550 and
4550-5200\AA, while the dM had four bins: 6600-7200,
7200-7800,
7800-8400, and 8400-9000\AA.  For each $\sim$600\AA\ bin, we shifted the wavelength regime by 10\AA\ 30
times towards lower wavelengths (blue) and 30 times towards longer
wavelengths (red) and calculated the RV using the
cross-correlation technique.  The
RV was taken to be the median value of all the
measurements and we report the uncertainty as the median absolute deviation.  The dM RVs are
much more reliable because of the many narrow molecular features found
in the dM spectrum, while the WD RVs have higher uncertainties due to their
broad hydrogen Balmer absorption lines.

\begin{figure}[!h]
\begin{center}
\includegraphics[trim=0cm 0.5cm 0cm 0cm, width=0.45\textwidth]{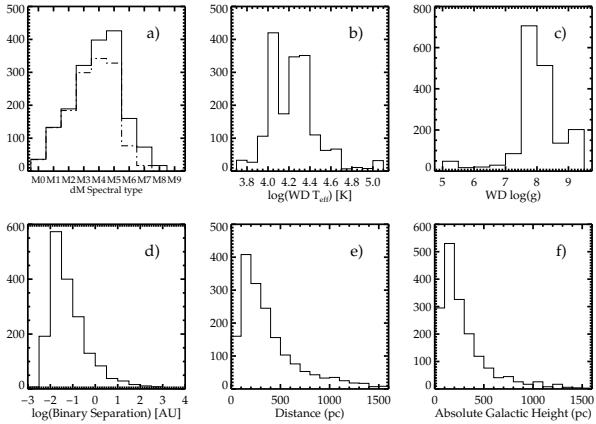}
\end{center}
\caption{Distributions of various properties of the sample:  a) the distribution 
of the dM spectral types of our sample (the black line is the entire sample, while the black dash-dotted line
  indicates the systems with high S/N that were used for the activity analyses);
 b) the distribution of the WD effective temperature as derived by our model
fitting procedure, which has a bimodal distribution at 10,000K and
19,000K; c) the distribution of derived WD $\log g$ values,
which peaks around a value of 8.0;  d) the distribution of 
the estimated binary projected physical separation in AU;  e) the distribution of the 
distances (pc) calculated using the spectroscopic parallax relation;
f) and distribution 
of calculated absolute distance above the Galactic plane (pc).}
\label{fig:all_hist}
\end{figure}

Some examples of the results from the iterative fitting process can be
found in Figure~\ref{fig:wddm_spectra}.  The top panel of the plot shows the original SDSS
spectrum in black, the derived dM flux in red (SDSS spectrum -- best-fit WD model), and the best-fit dM
template in orange.  The bottom panel is similar:  the original
SDSS spectrum is shown in black, the derived WD flux in light blue
(SDSS spectrum -- best-fit dM template), and the best-fit WD model in
dark blue.  Our procedure returned dM spectral type, WD
effective temperature ($T_{\rm eff}$), WD surface gravity ($\log g$), and WD and dM RV .  Histograms of these
parameters can be found in Figure~\ref{fig:all_hist} and the values can be found in
Table~\ref{tab:wddm1}.  We note that the WD parameters calculated through the above analysis are only estimates.  In the following section (Section~3.2) we use the resulting WD spectrum (from the WD$+$dM separation procedure described above) to perform a more detailed analysis to calculate WD parameters.  When available, calculated WD parameters reported from the analysis discussed in Section~3.2 and found Table~\ref{tab:wddm2} should be used.

\subsection{WD cooling analysis}
\begin{figure}[!ht]
\centering
\includegraphics[trim=0cm 0cm 0cm -0.5cm, width=0.47\textwidth]{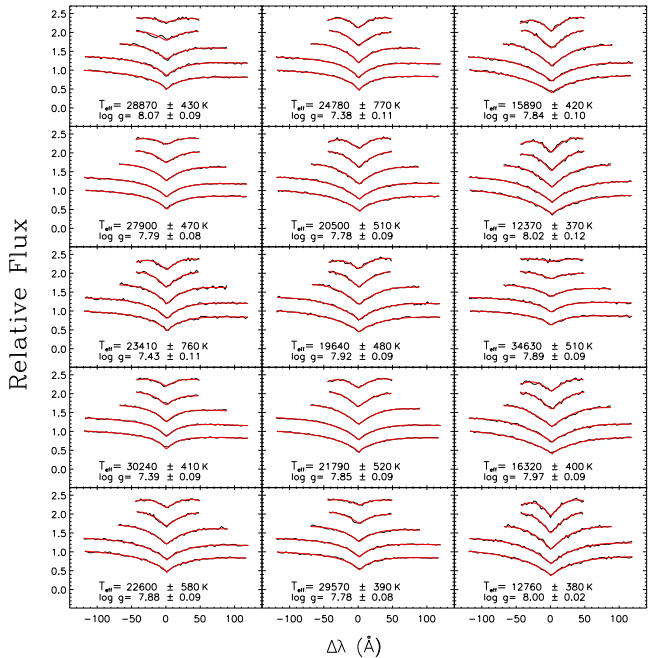}
\caption{Fits to the individual Balmer lines for the WDs for 15 separated WD in our sample. The line profiles correspond to H$\beta$ (\textit{bottom}) up to H$\epsilon$ or H$\zeta$ (\textit{top}). We have applied vertical shifts for clarity. The solid lines are the observed spectra and the red lines are the models that best fits the profile.}
\label{fig:balmer_fit}
\end{figure}

We derived the atmospheric parameters of the DA WDs ($T_{\rm eff}$ and 
$\log g$) by performing a fit of the observed Balmer lines to the WD 
models of D. Koester (private communication). The Balmer lines in such WD 
models were calculated with the modified Stark broadening profiles of 
\cite{Tremblay2009}, kindly made available by the authors. For the line fitting 
we used the code {\tt fitsb2}, which follows a procedure based on $\chi^{2}$ 
minimization \citep{Napiwotzki2004}. We considered the lines from H$\beta$ to 
H$\epsilon$ for the fitting procedure when S/N$<$30 and from H$\beta$ to 
H$\zeta$ when S/N$>$30. We did not include H$\alpha$ to minimize any effects from 
the M dwarf spectral subtraction technique (Section~3.1). As initial values, we 
considered the photometric temperature and a $\log g=8$. The photometric 
$T_{\rm eff}$ were obtained by fitting the SDSS synthetic {\it gri} magnitudes to WD 
synthetic photometry of \cite{Holberg2006}.  Since the SDSS {\it gri} magnitudes are likely
contaminated by the M dwarf component we derived synthetic {\it gri} magnitudes
from the spectrum of the WD component separated by the procedure outlined in
Section~3.1.  The entire WD fitting procedure is described 
in more detail in \cite{Garces2011} and \cite{Catalan2008a}. By using the photometric $T_{\rm eff}$ 
as initial value in the Balmer line fitting procedure we avoided possible 
degeneracies that result from the cool/hot solution \citep{Garces2011}.  In Figure~\ref{fig:balmer_fit} we show the fits to the 
Balmer line profiles for 15 DA WDs in our sample.

\begin{deluxetable*}{lrrrrrr}
   \tablewidth{0pt}
   \tablecolumns{7}
   \tabletypesize{\scriptsize}
   \tablecaption{WD$+$dM Parameters II: WD Cooling analysis parameters}
   \tablehead{\colhead{SDSS ID}                      &
	      \colhead{WD $T_{\rm eff}$ ($\sigma_{\textrm{WD} T_{\rm eff}}$)}			     &
	      \colhead{M$_{WD}$ ($\sigma_{\textrm{M}_{\textrm{WD}}}$)}			     &
	      \colhead{T$_{\textrm{cool}}$ ($\sigma_{\textrm{T}_{\textrm{cool}}}$)}\\
              \colhead{}                                &
	      \colhead{(K)}			     &
	      \colhead{(M$_{\sun}$)}			     &
	      \colhead{(Myr)}		     }
\startdata
SDSS J000152.10$+$000644.7 &     7598.0 (166.0) &      0.594 (0.0006) &   1150.300 (47.500)\\
SDSS J000356.93$-$050332.7 &    17617.0 (453.0) &      0.661 (0.0464) &    132.119 (14.636)\\
SDSS J000453.94$+$265420.4 &     9503.0 (285.0) &      0.599 (0.0080) &    651.520 (37.821)\\
SDSS J000504.92$+$243409.7 &    12916.0 (594.0) &      0.826 (0.0627) &    490.303 (78.049)\\
SDSS J000531.10$-$054343.2 &    13753.0 (1505.0) &      0.789 (0.1019) &    373.370 (105.280)\\
SDSS J000559.87$-$054416.1 &    29981.0 (523.0) &      0.767 (0.0532) &     16.788 (3.645)\\
SDSS J000651.91$+$284647.1 &    14558.0 (1609.0) &      0.511 (0.1057) &    160.609 (51.463)\\
SDSS J000711.41$+$283848.6 &    13401.0 (1854.0) &      0.732 (0.1254) &    350.238 (119.961)\\
SDSS J000829.92$+$273340.5 &    17068.0 (801.0) &      0.451 (0.0786) &     77.320 (16.291)\\
SDSS J 001247.18$+$001048.7 &    18843.0 (2922.0) &      0.993 (0.2610) &    289.300 (201.500)\\
\enddata
\label{tab:wddm2}
\end{deluxetable*}

According to \cite{Bergeron1992}, the atmospheres of DA WDs below $12000$K could 
be enriched in He while preserving their DA spectral type. This He is 
thought to be brought to the surface as a consequence of the development 
of a H convection zone. Depending on the efficiency of convection, the star 
could still show Balmer lines, instead of being converted into a non-DA 
white dwarf. The increase in He mimics the behavior of a high gravity, thus, 
the $\log g$ obtained from the spectroscopic fitting can not be trusted. 
For this reason when $T_{\rm eff}<$12000K we prefer to consider the 
photometric $T_{\rm eff}$ and the typical value for the surface gravity 
$\log g=8.0$.

Once we derived the $T_{\rm eff}$ and $\log g$ for each WD, its mass 
($M_{\rm WD}$) and cooling time ($t_{\rm cool}$) were obtained using 
appropriate cooling sequences. We have adopted the cooling tracks of 
\cite{Salaris2000}, which consider a carbon-oxygen (C/O) core WD (with a higher 
abundance of O at the central core) and a thick hydrogen envelope on top 
of a helium buffer, $q({\rm H}) = M_{\rm H}/M = 10^{-4}$ and $q({\rm He}) 
= M_{He}/M = 10^{-2}$. The resulting values for $M_{\rm WD}$ and $t_{\rm 
cool}$ are listed in Table~\ref{tab:wddm2}. The total age of a WD is comprised 
of its cooling time plus the lifetime of its progenitor star during the 
pre-white dwarf stage. If a white dwarf is isolated, or in a wide binary 
(no interaction between members), we can calculate the mass of the 
progenitor using an initial-final mass relationship  
\citep[e.g.,][]{Catalan2008b}, and considering stellar tracks, estimate the 
progenitor lifetime. However, in this study we are considering white dwarfs in 
close binaries that may have interacted in the past, transferring mass and altering the estimated progenitor properties. However, the cooling time 
of the WD can be used to constrain its total age by setting a 
lower bound. This lower limit is particularly important when the WD companion is an dM, since determining the ages of dMs is particularly challenging \citep{Soderblom2010}.  We derived $T_{\rm eff}$, $\log g$, $M_{\rm WD}$, and $t_{\rm cool}$
and their corresponding uncertainties for $904$ WDs in our WD$+$dM sample.
 The values are reported in Table~\ref{tab:wddm2}.

\subsection{WD$+$dM parameters}
In addition to the values presented in Tables~\ref{tab:wddm1} and~\ref{tab:wddm2}, we derived additional parameters that are given in Table~\ref{tab:wddm3}. In particular we derived the system
RV through space (km s$^{-1}$), the height above
the Galactic plane (pc), approximate distance to the system (pc), approximate
binary separation (AU) and orbital period (days) assuming circular, Keplarian orbits and
an edge-on inclination, activity state (as traced by H$\alpha$),
and magnetic activity strength (using
\lha).  In this section we discuss in detail on how we
calculated each of these parameters.

\subsubsection{System radial velocity}
In our sample, 1714 dMs and 1674 WDs had reliable RV
measurements, resulting in 1627 WD$+$dM pairs with RV measurements for both components.  In addition, we derived WD masses for 904 pairs as discussed in Section~3.2; for the rest of the WDs, we assigned a mass of 0.6 M$_{\odot}$ \citep[the average mass of 90\% of WDs in SDSS DR4;][]{Kepler2007}.  We inferred masses of the dMs from their spectral types  \citep{NLDS}.  Then using center-of-mass arguments in circular orbits, M$_{\textrm{dM}}v_{\textrm{dM}}$ = M$_{\textrm{WD}}v_{\textrm{WD}}$,
along with the measured relative RVs, we estimated the
system velocity of the pair.  The measured RVs are relative velocities
expressed as $v_{\textrm{wd}}^{'} = v_{\textrm{wd}} + v_{\textrm{sys}}$, where $v_{\textrm{wd}}^{'}$ is 
the measured RV of the WD, $v_{\textrm{sys}}$ is the
system velocity, and $v_{\textrm{wd}}$ is the absolute RV of the WD.
Similarly, we measured $v_{\textrm{dM}}^{'}$ for the dM:
$v_{\textrm{dM}}^{'} = v_{\textrm{dM}} + v_{\textrm{sys}}$.  Combining
these three equations, we solved for the system velocity:
\begin{equation}
v_{\textrm{sys}} = \left(v_{\textrm{wd}}^{'} +
\frac{M_{\textrm{dM}}}{M_{\textrm{WD}}}
v_{\textrm{dM}}^{'}\right) \left(1 +
  \frac{M_{\textrm{dM}}}{M_{\textrm{WD}}} \right)^{-1}.
\end{equation}

Absolute WD and dM RVs were calculated using the measured relative RVs and the derived system velocities. Absolute WD and dM RVs, system velocities, and their corresponding uncertainties are
reported in Table~\ref{tab:wddm3}; uncertainties were calculated using standard
error propagation.  We note that our reported system
velocities are lower limits due to the assumption of circular, edge-on, and location in the orbit.  In reality, most of our measured RVs have lower values than the true stellar orbital motions.
Thus, the system velocities are used for statistical relative
comparisons rather than exact measurements for individual pairs.

\subsubsection{Separations \& Periods}\label{Sec:Separations}
Without multi-epoch RV measurements, system parameters could not be
determined via full orbit solutions. Instead, we assumed the
single-epoch absolute RV of each component (or $V-r~\sin~i$) to be its orbital
velocity and calculated the projected linear separation for the binary
assuming a Keplerian circular orbit and edge-on inclination. The total
separation of a system was found by summing the individual component
separations from the center-of-mass of the system, which was
calculated by balancing centripetal and gravitational force.
Considering our calculated RVs are lower limits, this suggests that our
measured separations are likely upper limits.  When multiple RV
measurements were available (see Section~3.4), we constrained the limits on RVs and separations using the set of WD and dM measured RVs that had the maximum difference between them; this is where the RVs will be most representative of the orbital velocity.  While the projection effects
can be statistically corrected for ensembles, it is a steep function
of orbital separation and total system mass
\citep{DupuyLiu2011}. Applying the correction for individual systems
introduces large uncertainties and does not affect the relative
scalings; hence, we chose to not use the corrections. The estimated
projected linear separations (AU) and orbital periods (days) are
reported in Table~\ref{tab:wddm3}.

\subsubsection{Distances}
We used a spectroscopic parallax relation that was derived from data
in \cite{BochanskiPhD} to estimate the absolute
magnitudes of the dM as a function of spectral type and thus calculate the approximate distance to
the dM.  We estimated that the distances are accurate to 20\%.  Photometric parallax relations for dMs are more accurate
measures of distance \citep{W05,Davenport06,Bochanski2010,Bochanski2011}, however, due to the
photometric contamination by the close WD companion (in the $r$-band), the resulting
distances would have significant errors.  Calculated distances are
reported in Table~\ref{tab:wddm3} and a distribution of distances can be found in Figure~\ref{fig:all_hist}.

\begin{figure}[!ht]
\begin{center}$
\begin{array}{cc}
\includegraphics[trim=1cm 1.5cm 1cm 1cm,width=0.45\textwidth]{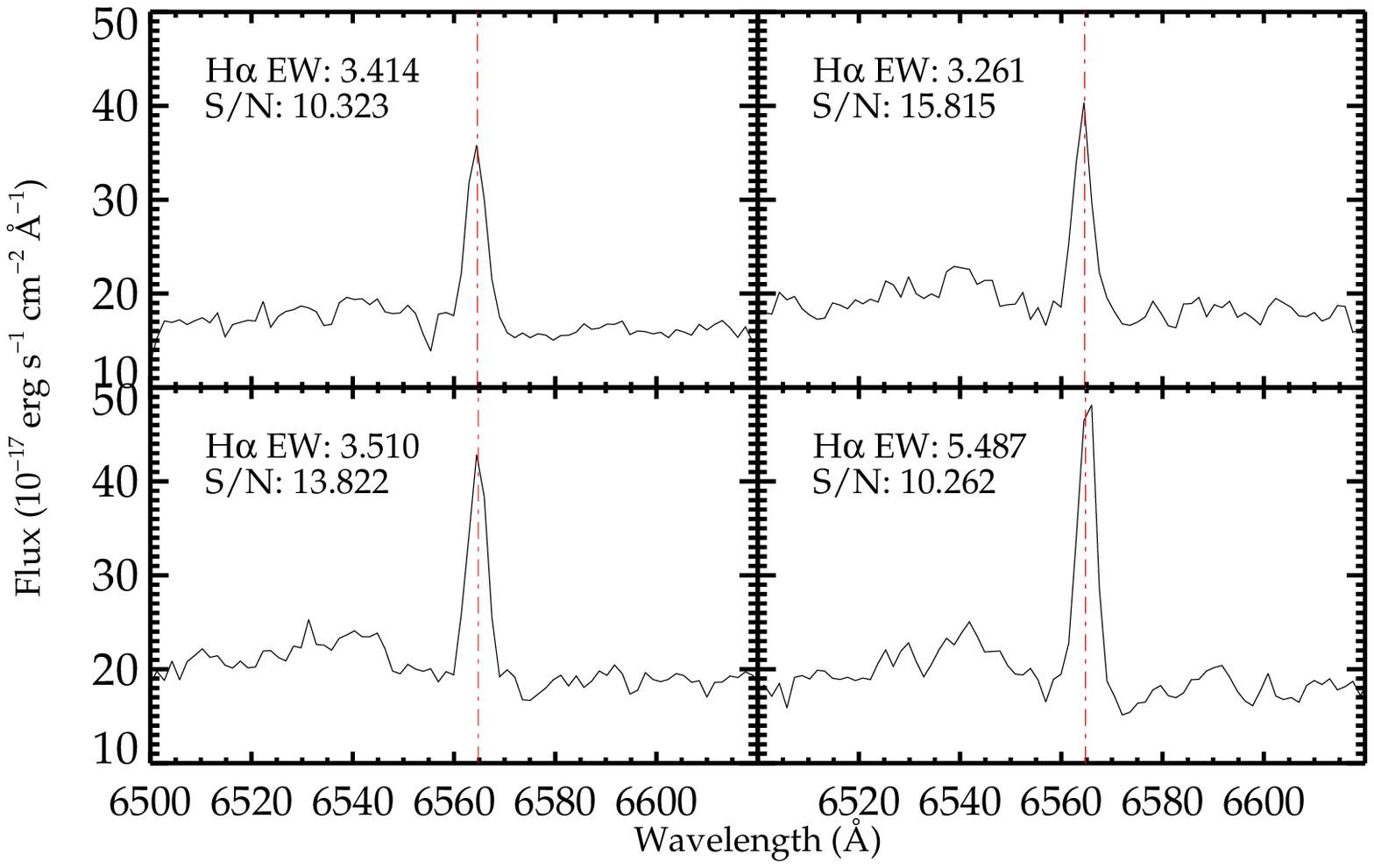} \\
\includegraphics[trim=1cm 2cm 1cm 1cm,width=0.45\textwidth]{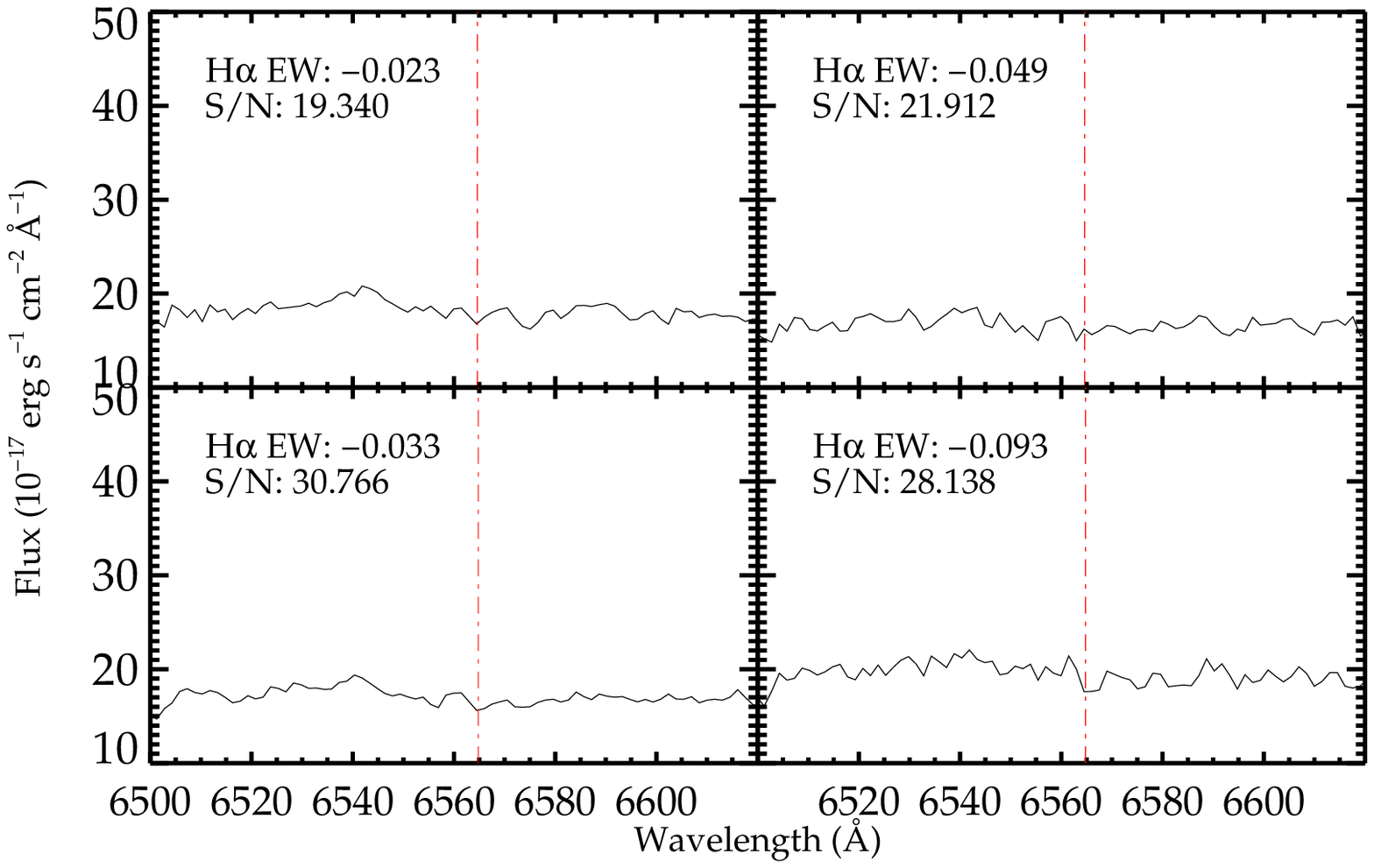}
\end{array}$
\end{center}
\caption{Examples active dMs (left panel) and inactive dMs (right panel) from our WD$+$dM sample.
  The red dot-dashed line is the location of the radial velocity
  corrected H$\alpha$ line as measured by our WD$+$dM separation
  procedure outlined in Section~3.1.  The H$\alpha$
  EW and S/N are printed on each panel.}
\label{fig:active_examples}
\end{figure}

\subsubsection{Vertical Distance from the Galactic plane}
Using the derived
distances from Section~3.3.3, the Galactic latitude and longitude, and by assuming the Sun is 15 pc above
the Galactic plane \citep{Cohen1995, Ng1997, Binney1997} and at a distance of 8.5 kpc from Galactic center \citep[e.g.,][]{Matsunaga2009}, we calculated the absolute height above
the Galactic plane for each pair using the following relation: 
\begin{equation}
Z_{Gal} = d
  \times \textrm{sin}(b - \textrm{atan}\left(\frac{z_{\sun}}{r_{\sun}}\right)*\textrm{cos}(l))
  + z_{\sun},
\end{equation}
where {\it Z$_{Gal}$} is the height above the Galactic plane, $l$
and $b$ are galactic latitude and longitude, $r_{\sun}$ is the
Sun's distance from Galactic center (pc), $z_{\sun}$ is the
Sun's height above the Galactic plane, and $d$ is the distance measured
from Section~3.3.3.  We used the height above the Galactic plane as a
statistical proxy
for the age of the binary system; younger stars are formed in the
plane of the Galaxy and over time are dynamically heated and perturbed
to heigher orbits (\citealt{W06}; W08). Distributions of calculated 
Galactic heights are reported in Figure~\ref{fig:all_hist} and the
values can be found in Table~\ref{tab:wddm3}.

\begin{deluxetable*}{lrrrrrrrrr}
   \tablewidth{0pt}
   \tablecolumns{10}
   \tabletypesize{\scriptsize}
   \tablecaption{WD$+$dM Parameters III: Additional parameters}
   \tablehead{\colhead{SDSS ID}                      &
	      \colhead{dM RV\tablenotemark{1}}       &
	      \colhead{WD RV\tablenotemark{2}}       &
	      \colhead{System RV\tablenotemark{3}}                    &
	      \colhead{$Z$}           &
	      \colhead{$d$}           &
	      \colhead{$s$\tablenotemark{4}}  &
	      \colhead{Period}                       &
	      \colhead{Active}                       &
	      \colhead{\lha~}                  	     \\
	      \colhead{}                             &
	      \colhead{(km s$^{-1}$)}                &
	      \colhead{(km s$^{-1}$)}	             &
	      \colhead{(km s$^{-1}$)}	             &
	      \colhead{(pc)\tablenotemark{5}}		             &
	      \colhead{(pc)\tablenotemark{6}} 		             &
	      \colhead{(AU)\tablenotemark{7}}		             &
	      \colhead{(days)}		             &
	      \colhead{}		             &
	      \colhead{}		             }
\startdata
SDSS J000109.42$+$255459.5 & -22.3 (2.6) & 208.3 (208.9) & 116.1 (125.7) & -223.6 & 410.3 & 0.08 & 8.5 & no & \nodata \\
SDSS J000152.10$+$000644.7 & \nodata & \nodata & \nodata & -894.6 & 1047.7 & \nodata & \nodata & yes & 1.04e-04 \\
SDSS J000250.65$-$045041.6 & -11.5 (8.4) & -36.6 (26.5) & -28.2 (17.7) & -275.0 & 320.3 & 8.54 & 9626.5 & yes & 1.63e-04 \\
SDSS J000356.93$-$050332.7 & -52.9 (6.3) & 73.6 (16.0) & 33.7 (14.5) & -121.8 & 150.7 & 0.40 & 93.8 & yes & 1.82e-04 \\
SDSS J000421.61$+$004341.5 & \nodata & \nodata & \nodata & -308.2 & 373.5 & \nodata & \nodata & no & \nodata \\
SDSS J000447.78$+$291140.9 & -18.0 (5.9) & -36.0 (25.9) & -27.0 (13.1) & -2070.1 & 3876.1 & 13.23 & 16066.7 & no & \nodata \\
SDSS J000453.94$+$265420.4 & -29.2 (11.4) & 174.5 (48.2) & 106.6 (33.9) & -192.8 & 364.3 & 0.13 & 18.0 & yes & 1.24e-04 \\
SDSS J000504.92$+$243409.7 & 41.2 (4.1) & 29.6 (14.5) & 32.7 (8.3) & -75.7 & 150.5 & 80.3 & 248091.09 & yes & 2.76e-04 \\
SDSS J000531.10$-$054343.2 & -24.0 (3.7) & -46.5 (85.9) & -40.8 (64.4) & -33.6 & 53.2 & 21.4 & 36401.82 & yes & 9.88e-05 \\
SDSS J000559.87$-$054416.1 & -9.6 (2.3) & 101.6 (35.7) & 60.3 (23.5) & -471.2 & 532.1 & 0.5 & 110.13 & yes & 8.82e-05
\enddata
\tablenotetext{1}{The absolute dM radial velocity corrected for the system velocity.  The corresponding dM RV uncertainties are reported in the parenthesis.}
\tablenotetext{2}{The absolute WD radial velocity corrected for the system velocity.  The corresponding WD RV uncertainties are reported in the parenthesis.}
\tablenotetext{3}{The system velocity as defined in Section~3.3.1.  The system RV uncertainties calculated from the dM and RV uncertainties using standard error propogation techniques are reported in the parenthesis.}
\tablenotetext{4}{Projected linear separation as defined in Section~3.3.2.  Recall that the
  separations presented in this paper are lower limits.}
\tablenotetext{5}{The absolute height above the Galactic plan calculated using the Equation 9}
\tablenotetext{6}{Distances were calculated using a spectroscopic
  parallax relation derived from data found in \cite{BochanskiPhD}.}
\tablenotetext{7}{The separations were calculated using the dM and WD RVs after being corrected by the system 
velocities.  We assumed that the measured absolute RVs of each component is the orbital velocity
as well as assuming edge-on circular Keplarian orbits.}
\label{tab:wddm3}
\end{deluxetable*}
\vspace{0.3in}
\subsubsection{Magnetic Activity}
We use the H$\alpha$ emission line as an indicator for magnetic activity in the dMs.  In order to calculate the magnetic activity strength, the line flux of H$\alpha$ was measured using a 
trapezoidal integration technique over 12\AA\ 
centered on the H$\alpha$ line \citep{W04}.  The continuum was taken as the
averaged flux 5\AA\ on either side of the line
region, the noise was the standard deviation of the flux over the
continuum range.  We calculated the \lha~of each dM using the methods described by \cite{Walkowicz04} and \cite{WestHawley2008}.  In \cite{Walkowicz04}, the authors derived a spectral type dependent scaling factor, $\chi$ (the ratio of continuum flux near H$\alpha$ over the bolometric flux), from a sample of nearby stars and 2MASS standards.  Multiplying the spectral type dependent $\chi$ values by the measured H$\alpha$ equivalent width (EW) resulted in \lha~independent of the dM distance.  \lha~is a measure of the flux output driven by magnetic activitiy via H$\alpha$ emission over the total bolometric luminosity of the star; this quantity is independent of the stellar continuum (unlike EW).  Magnetic activity strength as measured by \lha~is reported in Table~\ref{tab:wddm3}.

The magnetic activity state of the dMs was determined by inspecting the
H$\alpha$ line of each dM (WD subtracted) spectrum by-eye and
separating them into three main categories; 
active, inactive, or unable to classify due to low S/N spectra.  To
ensure the high quality of our classifications we only considered objects to
be active or inactive if the S/N near H$\alpha$ was greater than three.  Of our WD$+$dM
sample, we classified 741 as active and 496 as inactive -- 526 were
unclassified.  Four of our objects also have very broad H$\alpha$
emission features, possibly due to accretion
\citep[e.g.,][]{Hartigan1995}; we classified these separately and
excluded them from our magnetic activity analysis.  A few typical examples of active and inactive stars can
be found in Figure~\ref{fig:active_examples}. 
The activity states are reported in Table~\ref{tab:wddm3} as ``yes''
for active stars, ``no'' for inactive stars, and ``N/A'' for
unclassified stars.

\subsection{Multiple exposure RV analysis}
All SDSS spectra are composite
spectra made up of multiple exposures (at least three but as
many as 15).  The typical exposure times were roughly 900-1200
seconds and were often made in succession.  These individual exposures have been used in
previous studies to investigate M dwarf flare rates \citep{Hilton2010} and time variable
spectral features \citep[][]{Kruse2010,Bell2012}.  We used the multiple
exposure spectra in an effort to investigate time variability in the
derived radial velocity measurements in our close WD$+$dM binary pairs.
Many of our pairs are very closely separated, and thus may have
periods on the order of a few hours.  The single-epoch radial
velocities were calculated in a similar manner as discussed in Section~
3.1, except that the best-fit template for each spectrum was already
determined.  We also calculated \lha~for each individual exposure as
described in Section~3.3.5.  The number of exposures, the median change in radial
velocity ($\Delta$RV), the median error in the change of radial
velocity ($\sigma_{\Delta \textrm{RV}}$), and the time in between exposures of median RV change (hours) are reported in Table~\ref{tab:wddmrv}.  

Being able to measure multiple RVs during the orbit allowed us to put better constraints on the
separation of the WD$+$dM pairs as there was a greater chance of
catching the components at their maximum separation in one of the exposures (see Section~3.3.1).  Very close pairs
are apparent by the rapidly changing radial velocities over
the baseline of the consecutive observations.  In addition, these very close pairs
may be interesting for followup observations and may be used to investigate the variability of magnetic activity as
a function of orbital motion.
\vspace{0.3in}
\section{Results}
\subsection{Activity fraction vs. Galactic height}
\begin{figure}[!ht]
\begin{center}
\includegraphics[trim=1cm 1cm 1cm 0cm,width=0.45\textwidth]{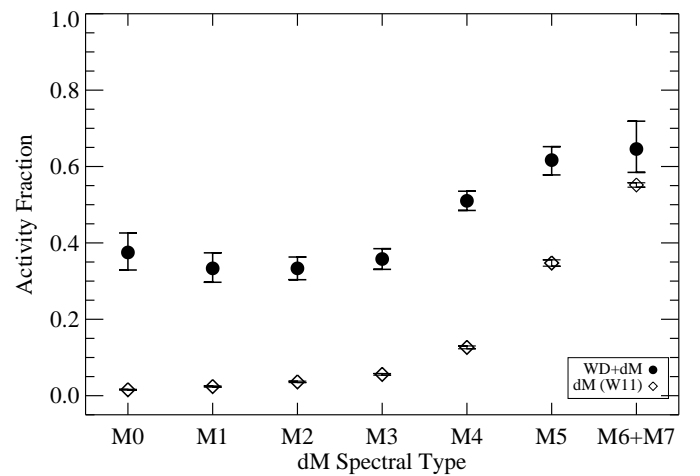}
\end{center}
\caption{Fraction of active stars vs. spectral type for our WD$+$dM
  sample (solid circles) and isolated field M dwarfs (open diamonds, W11).  The error bars were computed from the binomial distribution.  The M6 and M7 bins were grouped together
  to help increase the counting statistics for the later types. The
  fraction of active WD$+$dM is significantly higher than that of field
  M dwarfs.  Towards later spectral types we begin to see a decrease in the difference in activity fractions.}
\label{fig:act_spt}
\end{figure}

Figure~\ref{fig:act_spt} shows the
activity fraction in close WD$+$dM pairs (filled circles) as a function of spectral type
compared to the activity fractions in the dM field sample
(open diamonds, W11).  The uncertainties shown in
Figure~\ref{fig:act_spt} are calculated from the binomial distribution.  We see a noticeably higher activity fraction in the close WD$+$dM pairs than in the isolated field dMs across all spectral types.  However, towards later spectral types we see the disparity in the activity fractions decrease between the WD$+$dM and isolated dM populations.  The convergence of the two populations at late spectral types seems to indicate that having a close binary companion may not affect activity in late-type M dwarfs in the same manner that it does for the earlier-type dMs.

\begin{figure*}[!ht]
\begin{center}
\includegraphics[trim=1cm 0.5cm 1cm 1cm, width=0.8\textwidth]{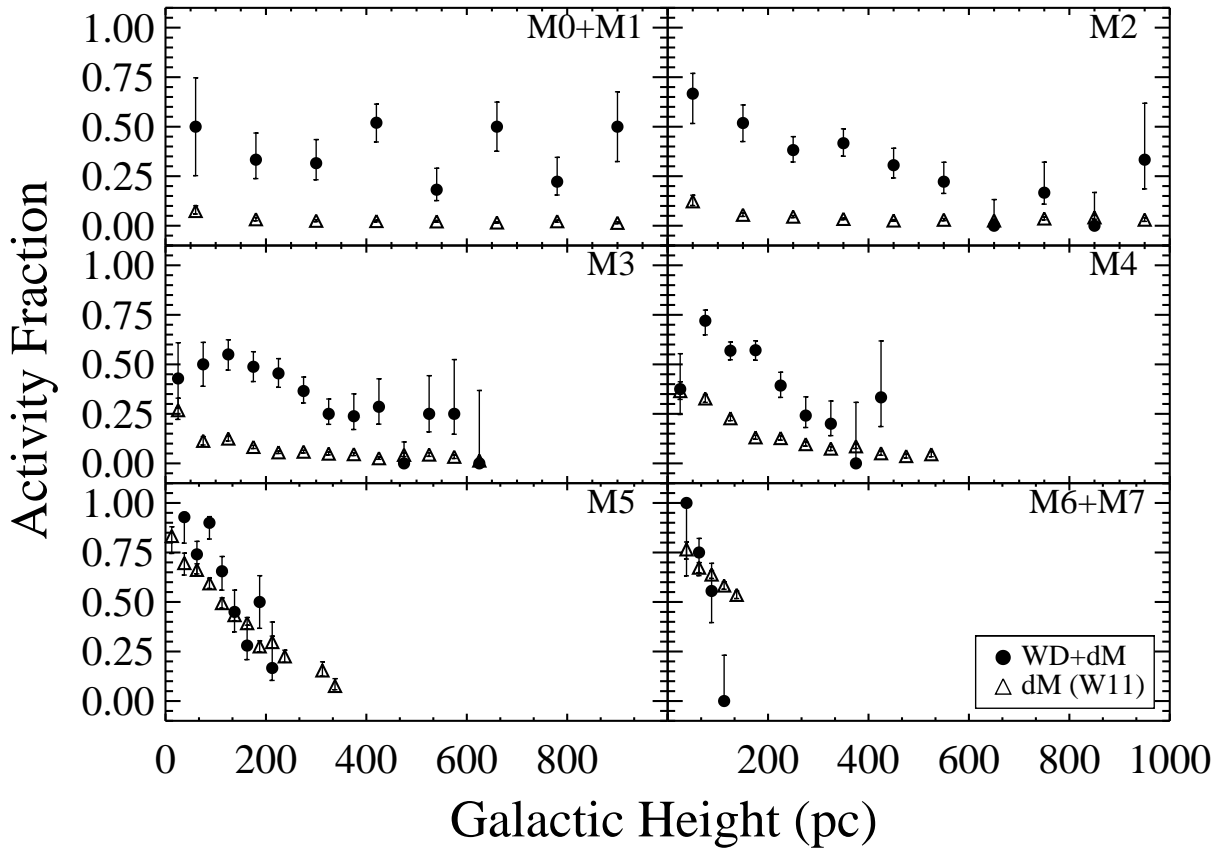}
\end{center}
\caption{Fraction of active stars as a function of absolute Galactic height (pc)
for each spectral type (proxy for age, W08).  
The filled circles are bins of WD$+$dM (a bin is required to have two or more stars) and the open triangles are bins of field dMs (W11).  In an effort to increase the binomial statistics in the least represented bins, we combined the M0 and M1 spectral types as well as the M6 and M7 spectral types. There are three main trends:
1) M dwarfs in WD$+$dM pairs are more active (up until M5) than their
single M dwarf counterparts;  2) whatever the mechanism driving the dynamo in fully
convective late-type dMs, it does not seem to be affected by a
close binary companion (or possibly rotation); and 3) we see elevated activity fractions close to the plane and the
presence of activity decreasing farther from the plane (age), indicating lifetime in early-type M
dwarfs with WD companions is most likely finite but lasts longer than field dwarfs.  
In late-type M dwarfs with WD companions the lifetimes
are comparable to the late-type field population activity lifetimes. }
\label{fig:act_z_spt}
\end{figure*}

Some of the morphology of Figure~\ref{fig:act_spt} is due to age effects (early-type M dwarfs have shorter active lifetimes than their late-type counterparts; W08) and the location of stars in the Galaxy (which is age dependent).  Figure~\ref{fig:act_z_spt} puts our sample in the proper Galactic context by showing the activity fraction as a function of absolute Galactic height (a proxy for age) for
each spectral type. The WD$+$dM (filled circles) sample is
binned by Galactic height and we chose our bin sizes to optimize the
number of stars per bin; the typical
bin (filled circles) has approximately 10 stars (with a minimum
of 2 stars). We used the same bin sizes for the field
M dwarfs (open triangles, W11).  Due to the relatively small
numbers of some spectral types, we
combined M0 with M1 and M6 with M7 for our analysis.  There are no M8
or M9 dwarfs with reliable activity parameters that were identified in
our analysis.  

Figure~\ref{fig:act_z_spt} shows that in comparison to isolated dMs (W11), the activity fraction is noticeably higher in almost all Galactic height bins until
a spectral type of M5, confirming the higher activity fraction trend seen in Figure~\ref{fig:act_spt}.  Also, at a spectral
type of M5, the disparity in activity fractions between field dM and
close WD$+$dM pairs decreases and continues to M6+M7 bins.  Due to the high uncertainties
the trend is not as clear in the latest type bins.  Coincidently, this change occurs roughly at the spectral types 
with interiors that are fully convective. In addition, the activity fraction as a function of Galactic height can be interpreted as evidence for activity lifetime.  By comparing the active fractions of dMs to simple 1D dynamical simulations, W08 showed that early-type dMs have short activity lifetimes ($\sim$0.6-2.0 Gyr for M0-M3, respectively) and that late-type dMs have longer activity lifetimes ($\sim$7.0-8.0 Gyr); there is a sharp increase of activity lifetimes at a spectral type of M4 ($\sim$4.5 Gyr), the location of the onset of full convection in dMs.  In Figure~\ref{fig:act_z_spt}, the low activity fraction in the isolated early-type dMs (W11) can be explained by the short activity lifetimes described by W08; the activity has already had enough time to ``turn off'' near the plane.  However, the isolated late-type dMs (W11) have high activity fractions close to the plane and continue to be active farther from the plane, indicating these stars remain active for longer.  When examining the close WD$+$dM population, Figure~\ref{fig:act_z_spt} shows a larger activity fraction (when compared to the field population) of early-type WD$+$dMs close to the plane and an elevated activity fraction decreasing steadily farther from the plane, whereas almost no early-type field dMs were active farther from the plane.   These results provide evidence for extended yet finite activity lifetimes for early-spectral type WD$+$dMs (M0-M3).  For the late-type dMs, Figure~\ref{fig:act_z_spt} shows that the close WD$+$dM activity fractions are similar to the late-type field dM population close to the plane and continue to be comparable at higher Galactic heights.  This similarity is interpreted as the late-type close WD$+$dMs and late-type field dMs having the same or similar activity lifetimes.  Coincidentally, M5 is just beyond the spectral type where the onset of full convection occurs in the interior of dMs.  

\subsection{Activity strength vs. Galactic height}
\begin{figure}[!ht]
\begin{center}
\includegraphics[trim=1cm 1cm 1cm 0cm,width=0.45\textwidth]{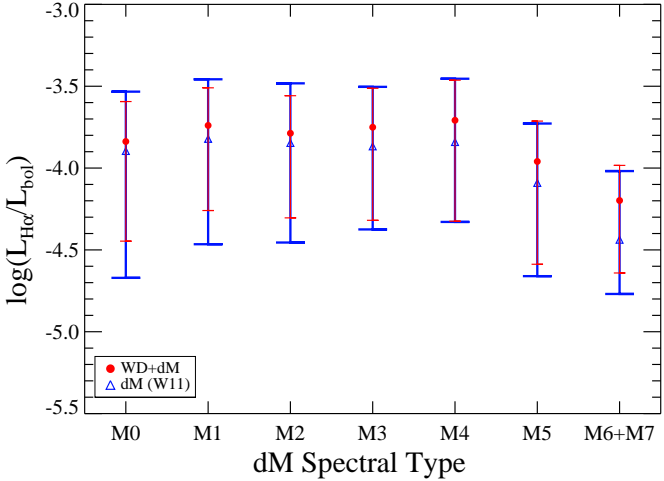}
\end{center}
\caption{Activity strength (as measured by \lha)
  vs. dM spectral type.  The data are binned by spectral type and
  the median value of \lha~is plotted.  The
error bars represent the 25th and 75th quartiles in that bin.  
  Red filled circles represent binned WD$+$dM and blue open
  triangles represent isolated dM from W11.  We see a decrease
in activity strength for field dwarfs at late types as well as a decrease in
strength for paired dwarfs.  The difference in activity strengths between the two populations in the M6+M7 bin indicates
that having a WD companion effects the activity strength but not the activity fraction
as was seen in Figure~\ref{fig:act_spt}.}
\label{fig:lhalbol_spt}
\end{figure}

Figures~\ref{fig:act_spt} and~\ref{fig:act_z_spt} indicate that WD$+$dM with early-type dMs are more
likely to be active than their field counterparts while late-type dMs in close WD$+$dM pairs show little
difference in their activity state compared to late-type field dMs .  Figure~\ref{fig:lhalbol_spt} examines the median activity strength (as measured by \lha) of the M dwarfs in WD$+$dM (filled red circles) in comparison to the median activity strength in field dwarfs (open blue triangles, W11).  The error bars represent the 25th and 75th quartiles of \lha~in each bin.  The activity strength in WD$+$dM pairs follows the same trend that we see in the field dM sample (W11); the activity strength is constant in early spectral types until a spectral type of M5, when the activity strength decreases.  The median activity strength in the WD$+$dM pairs is slightly larger than the median activity strength for the field dM population in every spectral type bin.  Similarly, the amount by which \lha~in WD$+$dM pairs is larger than field dMs remains constant ($\log(\lha)$$\sim+0.1$ dex) until a spectral type of M6+M7, where the difference in \lha~between WD$+$dM pairs and field dMs becomes noticeably larger ($\log(\lha)$$\sim+0.3$ dex).  This suggests that there is an increase in activity strength for later spectral types with close companions.  Unfortunately, we lack large numbers of late-type dMs (M8 and later) to explore this trend further.  To determine how similar the two (WD$+$dM and isolated dMs) populations are to one another, we performed a two-sided Kolmogorov-Smirnov test \citep[KS test,][]{Press1992} to determine whether the \lha~for all the spectral type bins between the WD$+$dM and isolated dM populations can be drawn from the same parent population.  We found an 87.1\% chance that the two populations are significantly different.  Figure~\ref{fig:lhalbol_spt} demonstrates that the activity strengths, along with the activity fractions, are consistently stronger for dMs with close WD companions.

\begin{figure}[!ht]
\begin{center}
\includegraphics[trim=1cm 0.5cm 1cm 0cm,width=0.45\textwidth]{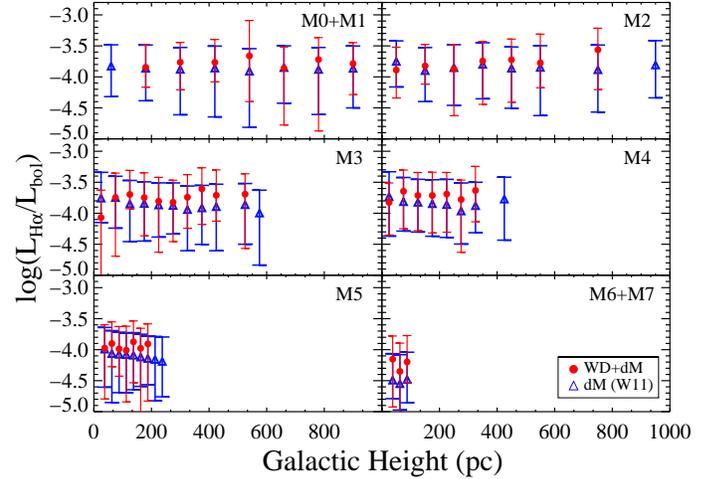}
\end{center}
\caption{Activity strength (as measured by \lha)
  as a function of Galactic Height for different dM spectral types.  
The filled red circles are the median \lha~value
of the bins of WD$+$dM, a bin is required to have 
two or more stars.  Isolated M dwarfs from W11 are 
represented in blue open triangles.  The error bars represent the 
25th and 75th quartiles of \lha~for each bin.  The WD$+$dM exhibit higher activity strengths than the field population across all spectral
types.  From spectral types from M3-M5 there is a slight decrease in the dM field populations \lha~with
increasing Galactic height, whereas, there appears to be no such decrease
with the WD$+$dM population.  Also, at later-types (M6+M7) there is a larger
difference in the \lha~values between the two populations than seen in earlier
spectral types.  A two-sided KS test found high 
significance that these two (M6+M7) populations are drawn from a separate parent
population.}
\label{fig:lhalbol_z_spt}
\end{figure}

Analogous to Figure~\ref{fig:act_z_spt},
Figure~\ref{fig:lhalbol_z_spt} investigates how the median activity
strength (traced by \lha)
behaves as a function of Galactic height and spectral
type.  We retain the same binning and formatting for Figure~\ref{fig:lhalbol_z_spt} that was used for Figure~\ref{fig:act_z_spt}.  Figure~\ref{fig:lhalbol_z_spt} shows that the activity strength of the dMs
in WD$+$dM pairs remains consistently higher than the activity strength of field
dMs with increasing Galactic height for spectral types M0-M4.  We
calculated the KS test for M0+M1, M2, M3, and M4 spectral type bins and found that there
is a 98.3, 66.2, 99.6, and 97.8\% chance, respectively, that the binary and field populations are
drawn from different parent populations.  At spectral
types $\ge$ M5, the activity strength in field dwarfs decreases with
increasing Galactic height (or age), a trend that is not as obvious in the
paired dMs.  Using the KS test we report a
99.8\% likelihood that the two M5 populations are
drawn from a different parent population.  Similar to 
M5, we see evidence that dMs in WD$+$dM with spectral types M6+M7 
had larger activity strengths than the field dwarf counterparts.  With
only three bins, we were unable to verify the decreasing activity
strength with increasing Galactic height as seen in the M5 bins.  
Since we only have three bins, we performed a KS test for the stars within each of the three
bins.  In order of increasing Galactic height, we found 98.9, 99.5, and 78.7\% significance that the two
populations are drawn from different populations.  Given the paucity of stars of our sample in the later spectral types,
we can only speculate that the activity strength appears to remain strong
in the later spectral types and also possibly with
increasing height (or age).  Figure~\ref{fig:lhalbol_z_spt} suggests that there may be a shift in
the behavior of activity strength (as a function of spectral type and
Galactic height) at spectral type M5, coinciding with a change in the structure of dM 
interiors.

\subsection{Activity \& Activity Strength vs. Binary Separation}
\begin{figure}[!ht]
\begin{center}
\includegraphics[trim=1cm 1cm 1cm 0cm,width=0.45\textwidth]{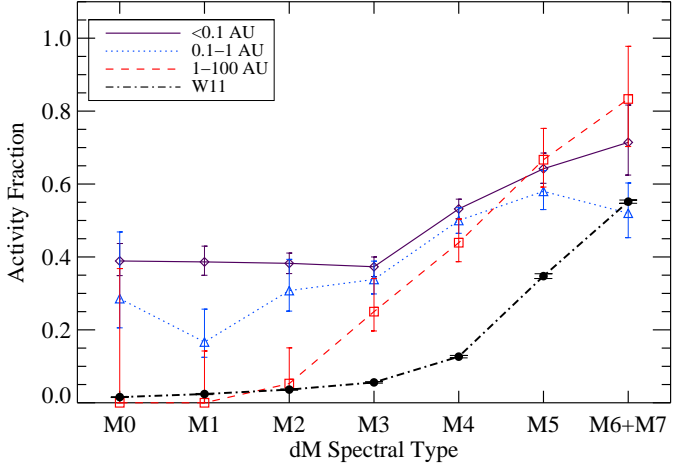}
\end{center}
\caption{Activity fraction for WD$+$dM versus spectral type grouped by projected,
linear separation: $\le$ 0.1 AU (puple solid line, diamonds), from 0.1--1 AU (blue dotted line, triangles), 1-100 AU (red dashed line, squares), and the W11 field dM population (black dash-dotted, solid circles).  In 
general, closer separated pairs tend to be more active than wider pairs, and even the widest pairs have elevated activity levels compared to the field population.  The narrowing of the activity fractions in the three separation bins towards later spectral types
is significant and may provide evidence that close companions no
longer effect the activity state of late-type dMs.}
\label{fig:sp_act_sep}
\end{figure}

In Figure \ref{fig:sp_act_sep}, we plot the fraction of active stars as a function of the dM
spectral type for three separate separation bins using our projected linear
separations (Section~3.3.2): $\le$ than 0.1 AU, 0.1-1 AU, and 1-100 AU.  Stars with separations $<$ 0.1 AU (purple solid line, diamonds) tend
to be more active than stars with separations between 0.1-1 AU (blue dotted line, triangles),
which are in turn more active than pairs with separations between
1-100 AU (red dashed line, squares).  We also include the W11 (black dash-dotted line, solid circles) isolated dM sample to compare activity behavior
between the most widely separated pairs and the field sample.  Figure~\ref{fig:sp_act_sep} demonstrates that at early spectral types, the widest pairs are unlikely to be active and have similar activity fractions as the field dM population.   Towards later spectral types, the wide pairs appear to be more active than isolated stars, however, this result is less robust due to the increasing uncertainty in the late spectral type bins for the widest pairs.  Figure \ref{fig:sp_act_sep} also shows that the difference in activity fractions between the
separation populations decreases with increasing spectral type.  We see a possible convergence of activity fraction in the later spectral type bins of
M6+M7.  Once again, this convergence appears to begin at the M3/M4
transition to fully convective interiors, which may indicate that a different
mechanism is responsible for generating the magnetic fields in late-type M dwarfs.

\begin{figure}[!ht]
\begin{center}
\includegraphics[trim=1cm 1cm 1cm 0cm,width=0.45\textwidth]{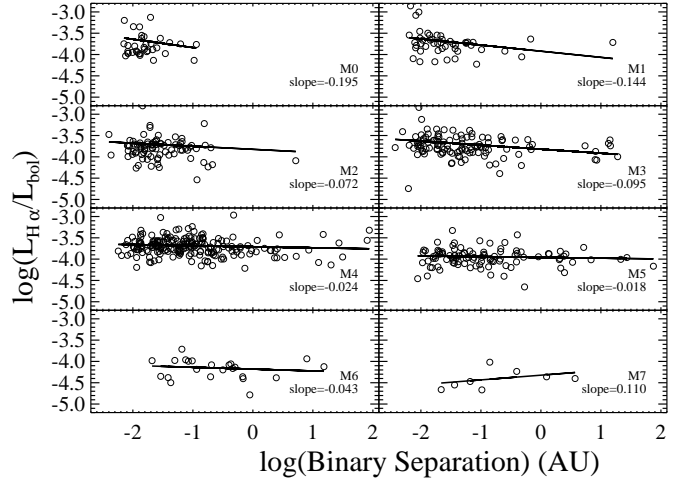}
\end{center}
\caption{$\log$(\lha) vs. $\log$(binary separation) for active dMs in close WD$+$dM pairs for each spectral type.  The solid line shows the linear least-squares fit for each distribution.  All of the spectral types exhibit a negative slope (with the exception of the M7, an inconsistency we attribute to the low numbers of pairs), implying higher magnetic activity strengths in the dM for smaller binary separations.  However, as discussed in Section~4.3, the linear least-squares fits are not statisticaly significant and the data are consistent with a horizontal line. }
\label{fig:lhalbol_sep}
\end{figure}

Figure \ref{fig:lhalbol_sep} shows the $\log$(\lha) vs. $\log$(binary separation) [AU] for all active pairs broken down by
spectral type.  For each spectral type bin we performed a linear
least-squares fit to the data; the slope of each fit can
be found on the corresponding panel in Figure \ref{fig:lhalbol_sep}.
We also performed a goodness-of-fit test (F Test) comparing the linear least-squares fit to a fit using a constant line through the data.  The significance values of the linear least-squares fits (as compared to a flat horizontal line) were 53.2, 11.3, 23.8, 1.9, 10.1, 53.5, 54.2, and 31.0\% for spectral types M0, M1, M2, M3, M4, M5, M6, and M7, respectively. 
Our results suggest that there is no conclusive evidence that the binary companion
increases the magnetic strength of the dM in a way that correlates
with binary separation.   We found no distinct change in
 $\log$(\lha) vs. $\log$(binary separation) across the M3/M4 fully convective boundary.
From Figures~\ref{fig:sp_act_sep} and~\ref{fig:lhalbol_sep}, we see that the proximity between the WD$+$dM
pairs does not appear to strongly affect the strength of the dM chromospheric activity.

\subsection{WD Cooling track ages}
\begin{figure}[!ht]
\begin{center}
\includegraphics[trim=1cm 1cm 1cm 1cm,width=0.48\textwidth]{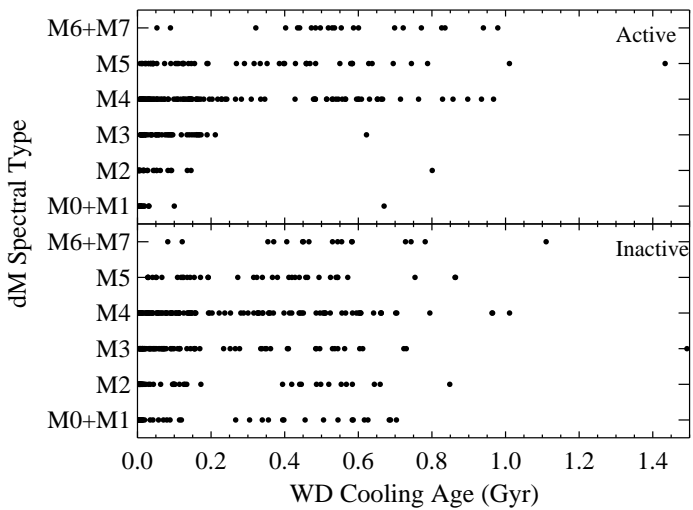}  
\end{center}
\caption{M dwarf spectral type vs. WD cooling age (Gyr) as a function
  of activity state. The top panel is the WD$+$dM pairs with active
  dMs and the bottom panel is the WD$+$dM pairs with inactive dMs.
  There is a dearth of old active early-type dMs in WD$+$dM pairs indicating that dM activity
  lifetimes are short-lived as is seen in the field dM population (W08).  By visual inspection there is no difference in the WD
  cooling ages spanned by the active and inactive late-type dMs.  A KS test for the spectral type bins M0+M1, M2, M3, M4,
  M5, and M6+7 found that there is a 2.5, 1.7, 2.5, 82.3, 46.9, and 100.0\% possibility, respectively, that the active and inactive populations are
  drawn from the same parent population.  Thus, for early-type WD$+$dM
  pairs, the active population is distinct from the inactive
  population, but there is no such difference for the late-type WD$+$dM populations.
  S06 conducted a similar study and also found a dearth of
  active early-type dMs in WD$+$dM pairs.}
\label{fig:tcool_spt}
\end{figure}

Using the analysis described in Section~3.2 we were able to derive physical properties for 904 of the WDs, including $T_{\textrm{eff}}$, mass, and cooling age (and their
associated uncertainties).  The WD masses
were used to calculate the system velocities as well as
the separations and periods of the pairs.  We used the WD cooling
ages as an alternative proxy for the age of the WD$+$dM systems.  The WD
cooling ages are accurate to a few 100 Myr for WDs with $T_{\textrm{eff}} < 25000$K and accurate to $\sim$20\% for WDs $>25000$K; the WD cooling ages do not take into account the progenitor lifetimes.  Therefore, the cooling ages are used in this study as a lower limit of the age and as a rough guide
to distinguish between old and young systems.  The progenitor
ages cannot be determined for these systems as many of these pairs
have close enough separations that mass transfer may have occurred;
a slight change in the WD mass will have a significant effect on the estimated
progenitor mass \citep[and its main sequence lifetime;  e.g.,][]{Catalan2008a}.  Figure~\ref{fig:tcool_spt}, shows the WD cooling ages (Gyr) as
a function of dM spectral type for the the active pairs (top panel) and the
inactive pairs (bottom panel).  WDs with $\log g~\textrm{errors} >0.4$ and S/N $<5$ were removed because of the large uncertainties in the calculated WD cooling ages.  Figure~\ref{fig:tcool_spt} shows that there is dearth of early-type WD$+$dM pairs
with active dMs in comparison to the inactive dMs. Past spectral types of M4, both
active and inactive pairs have a similar spread in WD cooling ages. 
We performed KS tests for spectral type bins M0+M1, M2, M3, M4, M5, and M6+7 and found that
there is a 2.5, 1.7, 2.5, 82.3, 46.9, and 100.0\% possibility, respectively, that the active and inactive
populations are drawn from the same parent population.  From the
results of the KS test, we conclude that early-type, active dMs in WD$+$dM
pairs tend to have younger WD companions than the inactive sample.  
We see less of a difference in the active and inactive populations in the
late-type dMs.  In the S06 study (in which they also used WD
cooling ages), WD$+$dM pairs with active dMs in well sampled spectral
type bins tended to be younger on average than WD$+$dM pairs with
inactive dMs.  Our results argue that only the early-type ($\le$ M3) active dMs
appear to be younger on average.

\begin{figure*}[!ht]
\begin{center}$
\begin{array}{cc}
\includegraphics[trim=1cm 1cm 0cm 1cm,width=0.45\textwidth]{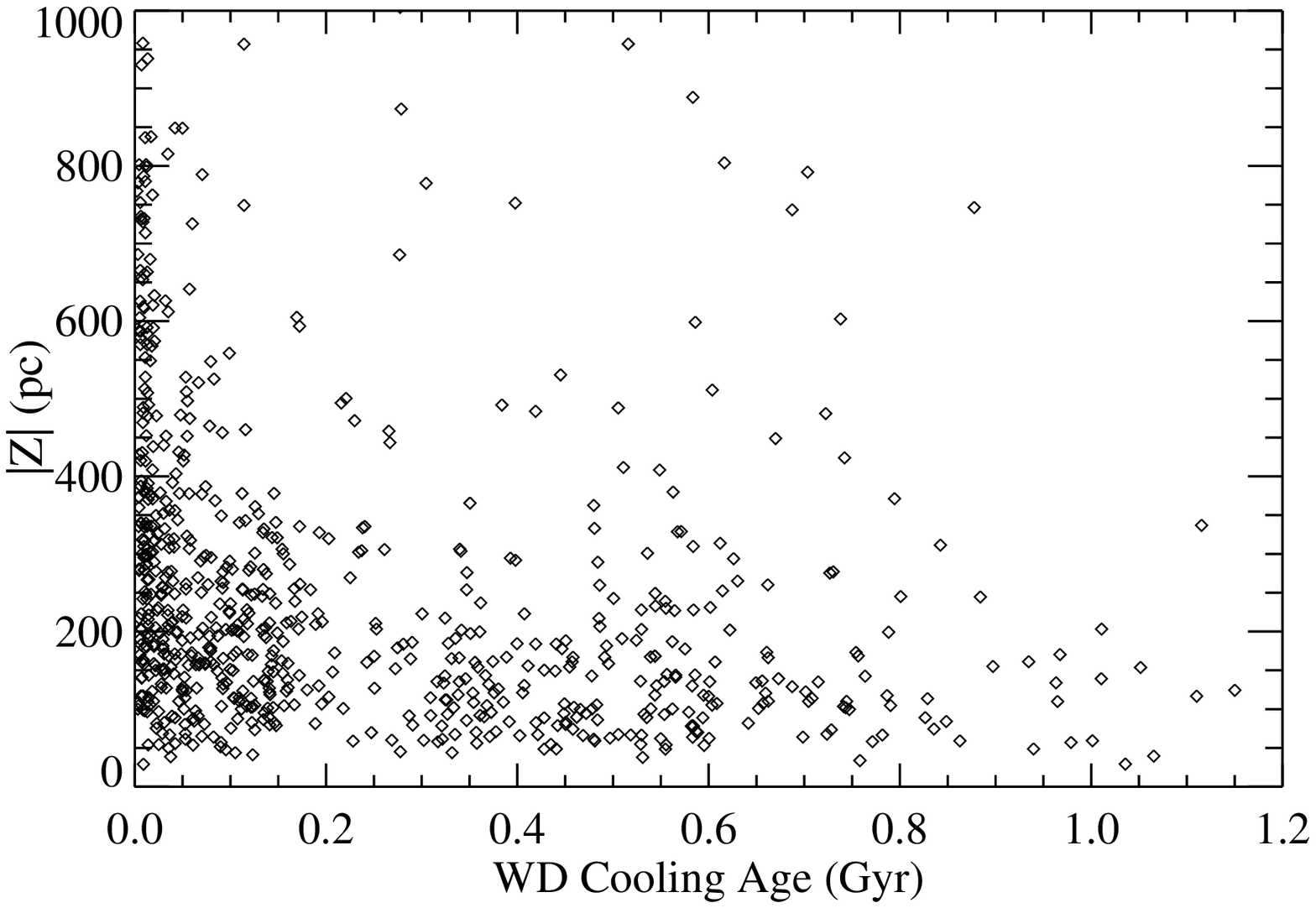} 
\includegraphics[trim=0cm 1cm 1cm 1cm,width=0.45\textwidth]{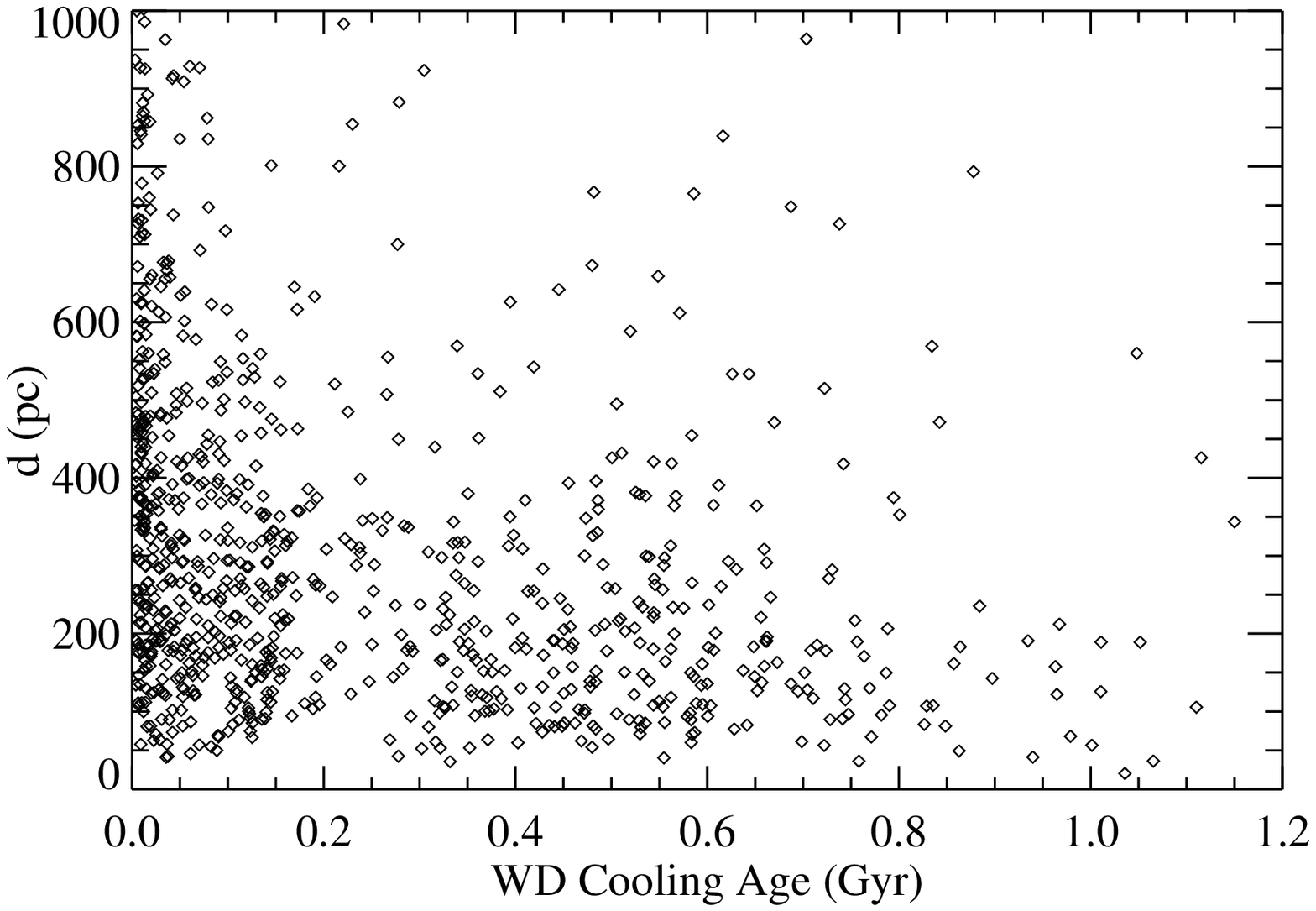} \\
\end{array}$
\end{center}
\caption{WD cooling age vs. absolute Galactic
  height (left panel) and WD cooling age vs. distance (right panel).
  The methods used to calculate WD cooling ages (Gyr), absolute Galactic
  Heights (pc), and distances (pc) can be found in Section~3.2, 3.3.4, 
  3.3.3, respectively.  The cluster of young WDs (WD cooling age
  $\sim$ 0.0) with a large spread in both Galactic height (left panel)
  and distance (right panel) indicates that our sample is
  biased towards WD$+$dM systems with young, bright WDs that we are
  capable of seeing at large distances.  Galactic height and distance are
  roughly equivalent properties due to the SDSS
  footprint, which focuses on the Northern Galactic cap.  We note that
  WD$+$dM pairs with distances $\le$ 300 pc span the entire range of WD cooling ages and are likely
  unaffected by the selection bias.}
\label{fig:tcool_z}
\end{figure*}

We note that there were three active, early-type WD$+$dM pairs that exhibited old WD cooling ages, which was incongruent with the rest of the sample.  We suspect that the two pairs in the M0/M1 and M2 spectral type bins were active due their very small separations ($\sim 0.05$AU; see Figure~\ref{fig:sp_act_sep}).  The WD$+$dM pair in the M3 spectral type bin did not have as small of a separation ($\sim 0.17$ AU), however, it is possible that an inactive dM was observed during a flare event \citep{Kowalski08}. 

In Figure~\ref{fig:tcool_z}, we plot the WD cooling age against the absolute
Galactic height (Figure~\ref{fig:tcool_z}; left panel) and
distance (Figure~\ref{fig:tcool_z}; right panel).  Both the WD cooling ages and Galactic heights are
proxies for the age of the binary system and we expected to
see a linear relationship in Figure~\ref{fig:tcool_z} (left panel), albeit
with some scatter.  However, instead of the expected linear
relationship we identified a potential selection bias; there
is an overdensity of systems with small WD cooling
ages that span a large range of Galactic heights.  Figure~\ref{fig:tcool_z} suggests that
our sample is biased towards systems that are younger; young WDs are intrinsically brighter and are easier to observe farther away.  Due to the
SDSS footprint (which focuses on the Northern Galactic cap), distance
and absolute Galactic height are roughly equivalent.  The similarity
between Galactic height and distance can be seen in similar trends
between the left and right panel of Figure~\ref{fig:tcool_z}.  Thus, the
distribution seen in Figure~\ref{fig:tcool_z} (left panel) is due to an oversampling
of the
bright , young WD components that are more easily observed and can
be found at farther distances.  We will
discuss the possible implications of this selection effect in more detail in Section~5

\subsection{Very close binary pairs}
\begin{deluxetable*}{lrrrrrrrr}
   \tablewidth{0pt}
   \tablecolumns{9}
   \tabletypesize{\scriptsize}
   \tablecaption{Very close WD$+$dM candidate pairs}
   \tablehead{\colhead{SDSS ID}                      &
	      \colhead{Plate}			     &
	      \colhead{MJD}			     &
	      \colhead{Fiber}			     &
	      \colhead{\# of}			     &
	      \colhead{$\Delta$RV}		     &
	      \colhead{$\sigma_{\Delta\textrm{RV}}$}	     &
	      \colhead{$\Delta$t} 	                        &	     
              \colhead{$s$}\\
	      \colhead{}			     &
	      \colhead{}			     &
	      \colhead{}			     &
	      \colhead{}			     &
	      \colhead{Exposures}		     &
	      \colhead{(km s$^{-1}$)\tablenotemark{1}}		     &
	      \colhead{(km s$^{-1}$)\tablenotemark{2}}		     &
	      \colhead{(hours)\tablenotemark{3}}                        &
              \colhead{(AU)\tablenotemark{4}} 
} 
\startdata
SDSS J001726.64$-$002451.2 & 687 & 52518 & 153 & 3 & 42.1 & 11.8 & 0.375 & 0.1564\\
SDSS J002157.91$-$110331.6 & 1913 & 53321 & 257 & 8 & 49.1 & 14.7 & 0.375 & 0.0227\\
SDSS J005245.12$-$005337.2 & 394 & 51812 & 96 & 5 & 42.1 & 25.5 & 0.281 & 0.0115\\
SDSS J015225.39$-$005808.6\tablenotemark{5} & 1504 & 52940 & 83 & 3 & 219.0 & 7.9 & 0.375 & 0.0758\\
SDSS J023938.04$+$273654.1 & 2444 & 54082 & 149 & 10 & 106.4 & 38.4 & 0.375 & 0.0109\\
SDSS J030138.24$+$050219.0\tablenotemark{6} & 2307 & 53710 & 140 & 3 & 64.9 & 27.7 & 0.469 & 0.1476\\
SDSS J030308.36$+$005444.1\tablenotemark{7} & 709 & 52205 & 524 & 4 & 330.9 & 13.4 & 0.469 & 0.0088\\
SDSS J073003.88$+$405450.1 & 2683 & 54153 & 224 & 6 & 26.3 & 7.5 & 0.375 & 0.0076\\
SDSS J073455.92$+$410537.5 & 2683 & 54153 & 507 & 6 & 31.1 & 10.3 & 0.375 & 0.0282\\
SDSS J075919.42$+$321948.5 & 890 & 52583 & 631 & 4 & 51.1 & 11.6 & 0.375 & 0.0955
\enddata
\tablenotetext{1}{We calculated the change in radial velocities from
  one exposure to another (most observations were taken in succession)
  .  $\Delta$RV is the median of the changes.}
\tablenotetext{2}{We report in this column the calculated the uncertainty in the reported
  $\Delta$RV using each exposures individual radial velocity measurement and uncertainty.  In order
  for the pair to be a very close WD$+$dM candidate $\Delta$RV $>$ $\sigma_{\Delta\textrm{RV}}$.}
\tablenotetext{3}{The time (hours) in between exposures for the
  reported $\Delta$RV.}
\tablenotetext{4}{Projected linear separation as defined in Section~3.3.4.  Recall that the
  separations presented in this paper are upper limits.}
\tablenotetext{5}{This system was identified as a close binary system in \cite{Nebot2011} and has a measured spectroscopic orbital period of 2.2h.}
\tablenotetext{6}{This system was identified as a close binary system in \cite{Nebot2011} and has a measured spectroscopic orbital period of 12.9h.}
\tablenotetext{7}{This system was identified as a eclipsing binary in \cite{Pyrzas2009} }
\label{tab:wddmrv}
\end{deluxetable*}

In our sample of 1756 pairs, 1347 had individual
exposures available in SDSS DR7 that were accessible to us from previous analyses (W11).  We found 37 objects that showed strong
evidence of large changes in RV in small time intervals.
The RVs and \lha~were calculated in the same way as was described
in Section~3.1 and Section~3.3.5, respectively.
The objects' primary identifiers, median change in RV,
median error in 
the change in radial velocity (constrained using the method described
in Section~3.1), and the median change in time in between exposures are reported in Table~\ref{tab:wddmrv}.  As well as
being interesting for studying the variability of WD$+$dM activity, these objects
are excellent candidates for studying very close binary pairs, mass accretion \citep[e.g.,][]{Hartigan1995}, eclipsing
binaries \citep[e.g.,][]{Pyrzas2009,Pyrzas2012,Parsons2012}, and pre-cataclysmic variables \citep[e.g.,][]{Schreiber2003,Tappert2009}.
We present these objects to the community as promising candidates for
follow-up observations.

\begin{figure}[!h]
\begin{center}$
\begin{array}{cc}
\includegraphics[width=0.36\textwidth, angle=90]{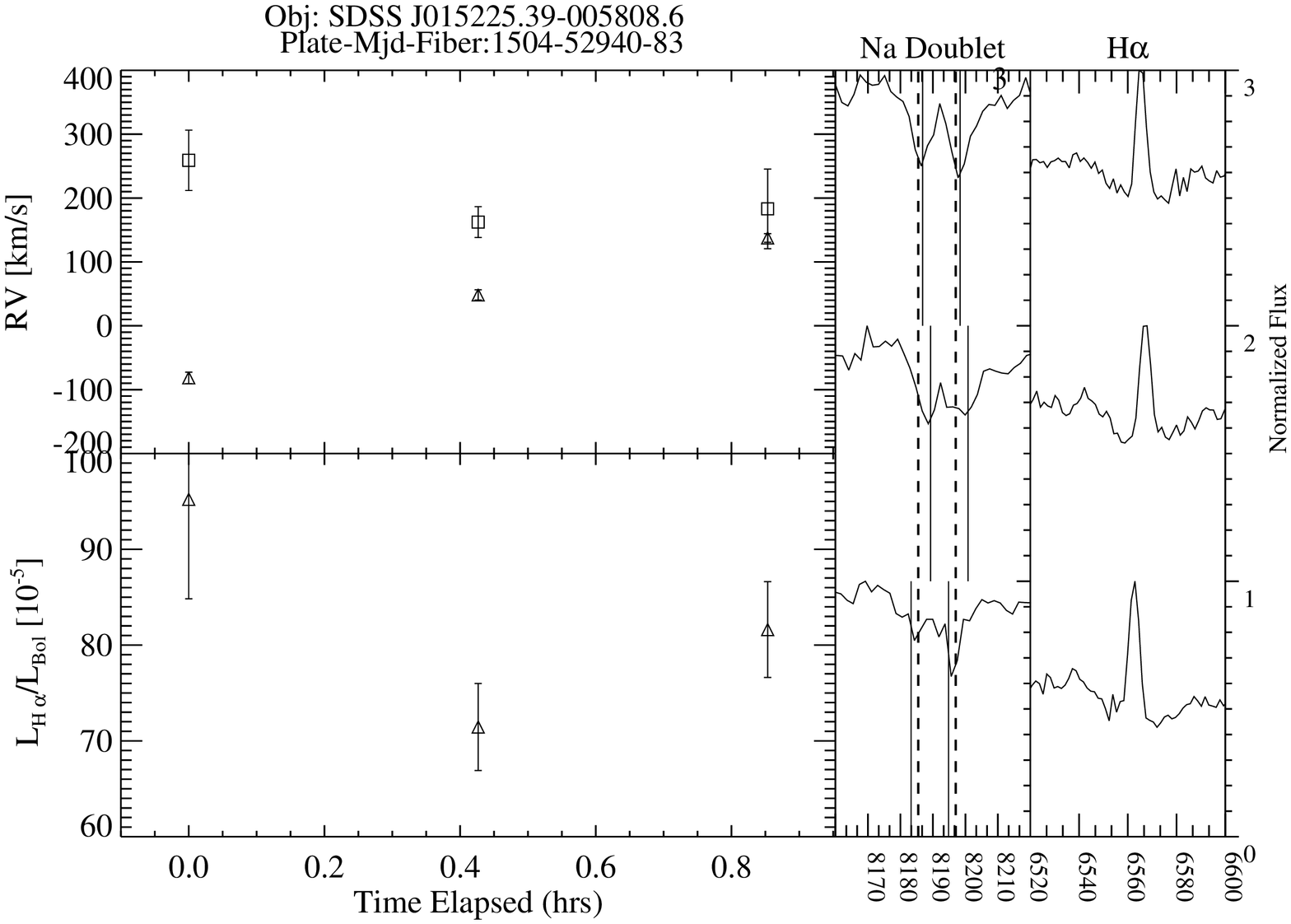} \\
\includegraphics[width=0.36\textwidth,angle=90]{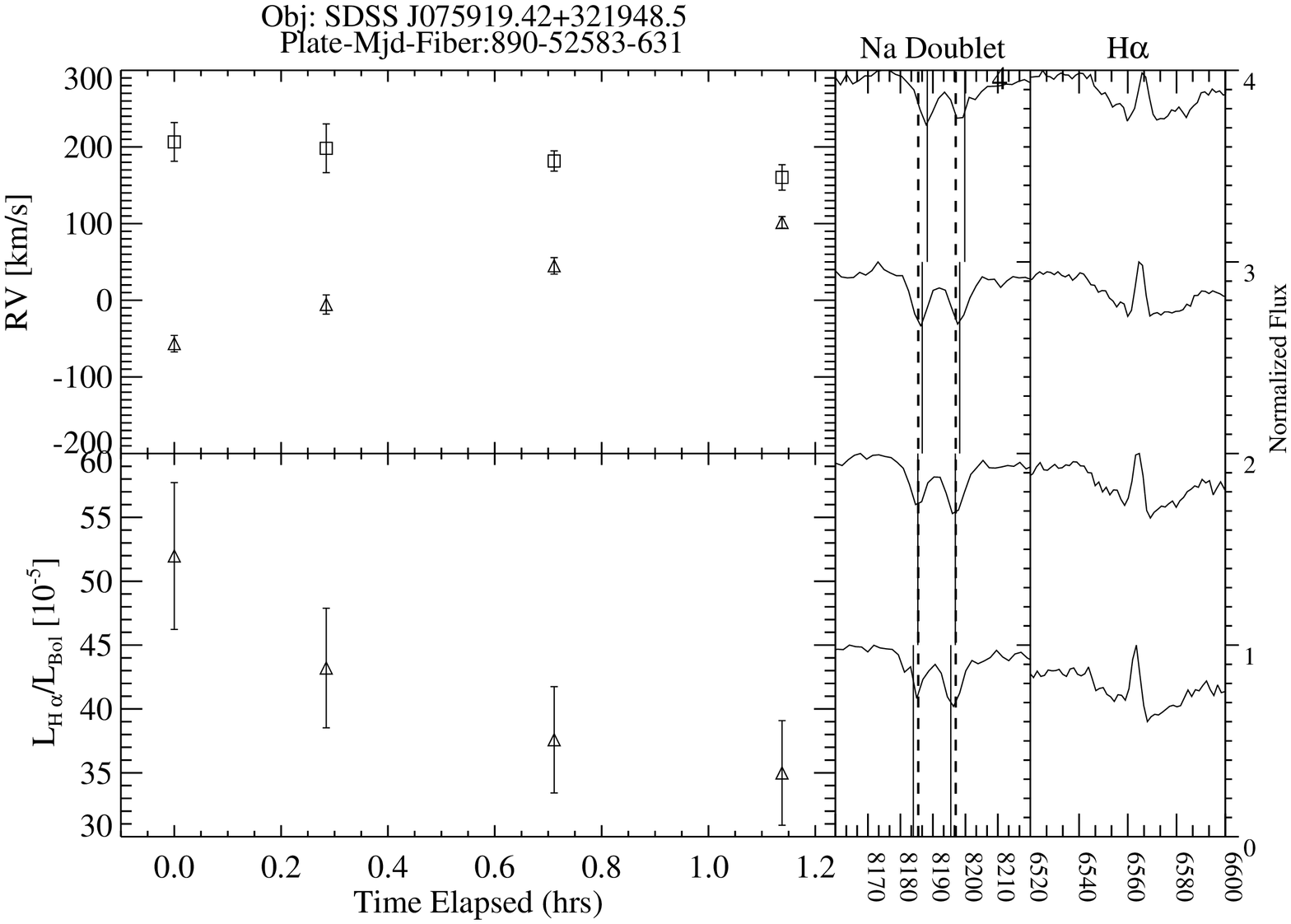} 
\end{array}$
\end{center}
\caption{Two examples of the very close WD$+$dM pairs.  Upper left panels: 
dM relative radial velocity vs. time elapsed since first 
observation. The black triangles are the dM radial velocity measurements while
the squares are the WD radial velocity measurements.  Lower left panel: 
\lha~for each exposure.  Middle panel:  The 8183\AA\ and 8194\AA\ Na absorption lines
from the dM for each exposure.  The vertical bars indicate the radial
velocity corrected location of the respective spectrum's Na I doublet lines, 
while the dashed lines indicate the rest wavelength of the Na absorption doublet
(8183\AA\ and 8194\AA\ ).  Right panel: The dM spectrum centered on the H$\alpha$
line.  The fluxes in the middle and right panel are normalized and
offset.}
\label{fig:multi_rv}
\end{figure}

Two of the most interesting of the 37 objects can be found in Figure~\ref{fig:multi_rv}.  For each pair, we plot the dM (triangles) and WD
(squares) relative radial velocities vs. time elapsed since the first observation (hours).  As discussed above, the WD RV calculations have larger uncertainties
associated with them due to the broad WD absorption lines.  Figure~\ref{fig:multi_rv} also
shows the \lha~vs. time elapsed in the bottom panel. Portions of each
spectra containing the 8183\AA\ and 8194\AA\ Na I absorption doublet and the
H$\alpha$ line are included to provide further evidence for our radial
velocity calculations.  The vertical bars indicate the predicted
location of the sodium doublet after being
corrected to the measured system radial velocity.  The fluxes
of the spectra are normalized and offset for visualization
purposes.

The top right panel
of Figure~\ref{fig:multi_rv} shows the dM RV for SDSS J015225.39$-$005808.6 
changing by $\sim$200 km s$^{-1}$ over the course of an hour along with a correponding change in the WD RV by $\sim$100 km s$^{-1}$.  Previous studies have identified SDSS J015225.39$-$005808.6 as a very close binary system with a spectroscopically determined orbital period of 2.2h \citep{Nebot2011}.  In the bottom plot of Figure~\ref{fig:multi_rv}, we see that the measured dM RV
for SDSS J075919.42$+$321948.5 increases by $\sim$200 km s$^{-1}$ over an hour while the WD RV decreases by the same amount.  We are likely seeing a significant component of the orbital motion in this pair in a short amount of time.  Previous studies have also identified SDSS J075919.42$+$321948.5 as a very close binary system and candidate eclipsing binary \citep{Nebot2011}. We suggest that the objects found in Table~\ref{tab:wddmrv} are promising candidates for followup studies pertaining mass accretion \citep[e.g.,][]{Hartigan1995} or eclisping binaries \citep[e.g.,][]{Pyrzas2009,Pyrzas2012,Nebot2011,Parsons2012}. 

\section{Discussion}
We investigated how the magnetic activity of dMs in close
WD$+$dM systems compared to that in the field population (W11).
In Figure~\ref{fig:act_spt} we found the activity fraction in WD$+$dM pairs was higher
than in isolated field dMs across all spectral types up to M7.  We
propose that the increased activity seen in WD$+$dM pairs is due to increased rotation in the paired dMs.   Many studies \citep[e.g.,][]{Skumanich1972, Pallavicini1981, Noyes1984,
  Delfosse1998, Pizzolato2003, Browning2010}  have correlated stellar activity with rotation in
main-sequence type stars to mid-to-late dM spectral types; the larger the
rotation velocities, the more activity one expects.  Thus, through
disk-disruption, tidal effects, and angular momentum exchange, we
suspect that the rotation of the dM in a close WD$+$dM binary system
is faster than that in a typical field dM. 
\cite{Meibom2007} argued that disk disruption during the
early stages of stellar dynamical evolution, which prevents or limits angular momentum dissipation and spin-down of the dM, is more likely the culprit than is tidal synchronization.  They found no correlation between stellar rotation and
orbital period or eccentricities among stars in the young ($\sim$150
Myr) M35 open cluster.  In addition, \cite{Meibom2007} offered the suggestion that
perhaps the observed faster rotation in close binaries could be due to
the varying amounts of intrinsic angular momentum present during the
formation and very early evolution of the closest binary stars versus
single stars.

Previous studies (e.g., S06)  also found
that early-type dMs ($\le$ M4) with a close WD companion have
higher activity fractions than isolated field dwarfs.  Due to the small number of stars with spectral types $\ge$ M5, S06 were unable to
investigate the activity fractions of late-type WD$+$dMs in comparison with field
dwarfs from W08.  Our results for the early-type dMs agree with
the results of S06 and our statistics are
better with over 70,000 field dMs from W11 as well as 1237
WD$+$dMs with reliable activity properties from our sample (compared to 661 WD$+$dMs with activity properties from S06).  The S06 study also lacked
sufficient numbers to explore the activity fractions in spectral
types $\ge$ M5 with any statistical significance.  We have extended
the activity fraction analysis in close WD$+$dM pairs out to spectral types $\sim$
M7.  We show with statistical significance that the trend of higher
activity fractions for dMs in WD$+$dM pairs in comparison to field dMs
extends at least to spectral types of M7 (Figure~\ref{fig:act_spt}).  

An interesting feature worth noting from Figure~\ref{fig:act_spt} is the convergence
between the active fractions of binary and single dM populations towards later spectral types.
Figure~\ref{fig:act_spt} shows that the activity fraction in later spectral type dMs with
close binary companions is not increasing as quickly as field dMs seen
by W11.  If the two populations truly converge at spectral
types $\ge$ M7, as Figure~\ref{fig:act_spt} suggests, then activity of late-type
dMs in close WD$+$dM pairs increases by a smaller amount in comparison
with field dMs.  Due to the long lifetimes of M
dwarfs and the fact that their activity state is correlated with their
location in the Galaxy, it is more appropriate to analyze their
activity fraction within the proper Galactic context (\citealt{Kowalski08,Hilton2010};~W08,~W11).  Figure~\ref{fig:act_z_spt} confirms that dMs in close WD$+$dM pairs have an increased likelihood of being active.  We assume that the main distinguishing property
between dMs with close WD companions and field dMs is that the former have more rapid
rotation (magnetic locking, disk disruption,
tidal effects, or angular momentum exchange) than we might
expect.  This implies that the mechanism driving magnetic field generation in
late-type dMs may be less sensitive to rotation than the
$\alpha$-$\Omega$-dynamo thought to be responsible for magnetic fields in
stars with masses greater than $\sim$ 0.33 M$_{\sun}$.  Previous theoretical studies have developed approaches for
generating magnetic fields in fully convective stars that are correlated with rotation \citep{Chabrier2006,Browning2008}.  Observationally, a few studies have suggested that the correlation between rotation and activity in fully convective stars
may not be as strong as theory predicts \citep{WestBasri2009,Reiners2012}.  Both \cite{WestBasri2009} and \cite{Reiners2012}
provide observations moderately rotating ($\ge$ 3.5 km s$^{-1}$), late-type dMs without any observable
magnetic activity.   While some of these previous results may be spurious due to low S/N spectra or other systematics, they offer a tantalizing possibility about the limited role of rotation in generating magnetic activity in late-type dMs. 

Figure~\ref{fig:act_z_spt} also examines the time evolution of magnetic activity, where Galactic height was used as a proxy for the approximate age of the system.  We assume
all of the stars are formed close to the mid-plane of the Galaxy and over their
lifetimes they are systematically able to move
further from the plane due to dynamical interactions. Figure~\ref{fig:act_z_spt}  shows that
the activity fraction is significantly larger across all the Galactic
height bins, indicating that not only are the early-type WD$+$dMs more
likely to be active than the field population, but also have longer
activity lifetimes (see W08).  While the activity lifetimes in early-type WD$+$dMs may indeed be longer lived, Figure~\ref{fig:tcool_spt} (WD cooling ages as a
function of spectral type for the active and inactive dMs) shows that there is
a dearth of older active early-type dMs.  The cooling age analysis in addition to the fact that not all early-type dM are active indicates that early-type dMs in close
WD$+$dM binary pairs still have finite active lifetimes (also seen in S06). This result is expected and agrees with the idea
that dMs in WD$+$dM pairs are sped-up (with respect to field dMs) and
still contain a solar-like $\alpha$-$\Omega$ dynamo.  However, for the
late-type WD$+$dMs ($\ge$ M5), the difference in the activity fractions
becomes less obvious.  In the M5 panel of Figure~\ref{fig:act_z_spt}, the WD$+$dM pairs
trace the activity fraction of the field dMs very well, implying that
 the presence of a close WD companion no longer has an appreciable
 effect on the activity of dMs with spectral types $\ge$ M5.  Not only is the
 activity fraction consistent between the two populations at spectral
 type M5, but this similarity implies that the activity lifetimes are comparable between the two
 populations.  This is corroborated by the WD cooling age analysis (Figure~\ref{fig:tcool_spt}), where
 the active and inactive late-type WD$+$dM pairs appear to be drawn
 from the same parent distribution with higher confidence (82.3, 46.9, and
 100.0\% for spectral type bins M4, M5, and M6+7) compared to the early-type WD$+$dM pairs (2.5, 1.7, and 2.5\% for spectral type bines M0+M1, M2, and M3).  With our assumption that dMs in close WD$+$dM pairs are
 rotating faster than the field population, this result suggests
 that rotation is not as important in the activity state of late-type
 paired dMs.  Evidence for this trend continues into the M6+M7
 spectral types.  Figure~\ref{fig:act_z_spt} suggests that whatever the driving
mechanism for magnetic field generation in late-type dMs, it may
not be affected by a close binary
companion, or even more dramatically, rotation.  The absence of enhanced activity in late-type WD$+$dM systems can also
be explained if the rotation-activity relation saturates at lower
levels than in the early-type dMs. In this case, late-type dMs in close WD$+$dM pairs are rotating faster than their single
counterparts. However, once the threshold is reached, increased 
rotation will not cause an increase in magnetic activity \citep[e.g.,][]{Pizzolato2003,Browning2010}.

\begin{figure}[!ht]
\begin{center}
\includegraphics[width=0.5\textwidth]{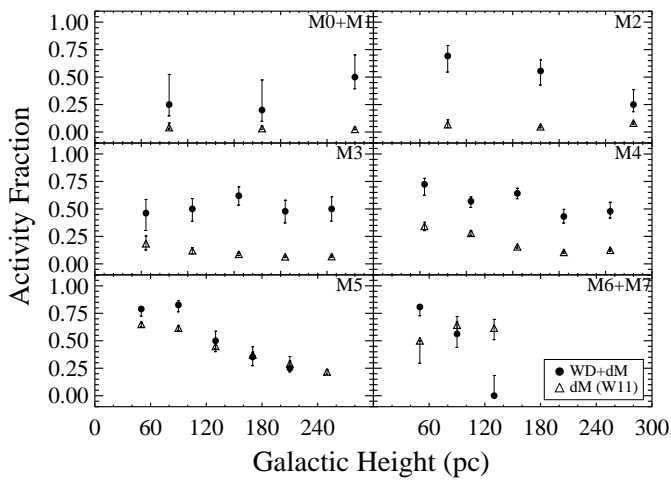}  
\end{center}
\caption{Fraction of active stars as a function of absolute Galactic height (pc)
for each spectral type for stars with distances $<$ 300 pc. The distance selection was introduced to test the severity of the selection bias identified
by Figure~\ref{fig:tcool_z}.
In general, the lack of distant, young systems does not affect the activity fractions and we conclude that the selection bias
does not significantly change our results.}
\label{fig:act_z_spt_bias}
\end{figure}

Figure~\ref{fig:tcool_z} shows that our sample  may be biased
towards younger systems, where both components are intrinsically
brighter and easier to observe at greater distances.  We therefore, may be oversampling young WD$+$dM pairs, which (at least for early-type
dMs) tend to be more active as seen in Figures~\ref{fig:act_z_spt} and
\ref{fig:tcool_spt}.  Bins of higher Galactic Height (given the SDSS lines-of-sight are concentrated near the Northern Galactic Cap) will then be
preferentially oversampled by active WD$+$dM stars simply because they
are brighter and easier to observe, tending to falsely inflate the activity
fractions in the more distant Galactic Height bins.  We examined this potential selection effect by remaking Figure~\ref{fig:act_z_spt} but only including WD$+$dM pairs with
distances $\le$300 pc, where we would be sensitive to WD$+$dM pairs of all WD cooling ages (Figure~\ref{fig:act_z_spt_bias}).  We chose to cut on distances $\le$300 pc because of the smooth sample
distribution seen below this cutoff distance (see
Figure~\ref{fig:tcool_z}; right panel).  The distance cut reduced our
sample to 892 WD$+$dM pairs, which has the consequence of enlarging the uncertainties in many bins. Figure~\ref{fig:act_z_spt_bias}
shows similar behavior to Figure~\ref{fig:act_z_spt}, namely
increased activity fraction in the early-type dMs in WD$+$dM pairs in
comparison to the field population (with the exception of one bin in
M0+M1 which is likely due to small number statistics) and similar activity fractions
between the late-type WD$+$dM and field dM populations.  The trend of decreasing
activity with Galactic height is somewhat lost in the M2 and M3
spectral type bins due to the small numbers and the fact we are sampling a smaller
range of Galactic heights.  We conclude that having a preferential selection bias
towards brighter and therefore younger systems at large distances does
not make a qualitative difference in our results.

Figures~\ref{fig:act_spt} and \ref{fig:act_z_spt} explored the behavior of the activity state (and activity lifetimes)
of dMs with close WD companions.  We found evidence that the presence
of close companions increases the likelihood that a star is active and extends the lifetimes of
early-type dMs.  However, for late-type dMs, close companions did not
have much of an effect on the activity and activity lifetimes of
late-type dMs.  We used Figure~\ref{fig:lhalbol_spt} to make an analogous study, except exploring activity strength (via \lha) of close WD$+$dM
pairs in comparison to the field dM population.  In Figure~\ref{fig:lhalbol_spt}, the median activity strength of the active WD$+$dM
population is slightly higher across all spectral types in comparison
to the field dM population. Also, \lha~is shown to remain relatively constant at approximately \lha~= 10$^{-3.7}$ from
spectral types M0-M4; a trend that is consistent with the field dM
population (\citealt{Burgasser2002,Gizis2002}; W04).  Towards the later spectral types ($\ge$ M5), there is a
decrease in activity in both populations (also seen in previous studies,~\citealt{Gizis2002,W04,W06}; W11).  In the M6+M7 spectral type bin
we see a distinctly larger difference in \lha~between the two
populations.  The question is whether or not the M6+M7 result is real or
simply due to a small number of stars in the M6+M7 spectral type bin.  If it is a
real result, then it could possibly provide some interesting insight
into the magnetic field generation in stars with fully convective envelopes.  Figures~\ref{fig:act_spt} and \ref{fig:act_z_spt} show that the activity state in later spectral types does not seem to be
affected by having a close binary companion,
however, the strength of activity might be
amplified by the WD.  If having a close companion imparts a
larger rotation to the dM, then we could be seeing that rotation in
later spectral types influences the strength in activity, but has less
of an effect on the likelihood of activity.  In other words, rotation may
play a limited role establishing magnetic activity but may have an
effect on the strength of the chromospheric heating.

Figure~\ref{fig:lhalbol_z_spt} puts strength for each spectral type bin in the proper Galactic context.  Figure~\ref{fig:lhalbol_z_spt} shows that the \lha~for M0+M1 and M2 dwarfs appears to be mostly flat
for both the WD$+$dM and field populations for most Galactic heights, but that the field dMs at spectral types of M3 and M4 show a slight decrease in
activity with increasing Galactic height.  There is no clear trend of decreasing \lha~with Galactic height seen
in spectral types M3-M4 for the WD$+$dM population. At spectral
types of M5 there is a large decrease in overall activity strength
 in both populations, and a distinct decline in \lha~with
increasing Galactic height in the field population, a trend that is indicative of activity strength decreasing with age.  There is no apparent  trend in the WD$+$dM population that shows a
decrease in \lha~with Galactic height as seen in the field dMs.
While there are only a few objects in the M6+M7 bin, it is curious that there is a fairly large difference in activity strength between the two
populations, as was identified in Figure~\ref{fig:lhalbol_spt}.  We reiterate that this could be
evidence for close companions affecting the activity strength in dMs
but not increasing the likelihood of activity or the lifetime of
activity (Figure~\ref{fig:act_z_spt}).

We have shown convincing evidence that close companions have an
effect on both the activity state and activity strength in dMs.  As
was discussed in Section~3.3.2, we have an approximate measure for the
separation for which we can investigate the effect of separation on
the activity of dMs.  Figure~\ref{fig:sp_act_sep} is presented in the
same way as Figure~\ref{fig:act_spt}, 
where the activity fraction for the WD$+$dM population is shown as a function of spectral
type.  The active WD$+$dM population is further split into three
separate separation bins, those less than 0.1 AU, 0.1-1 AU, and 1-100
AU.  These separations bins were chosen to ensure good statistics in each
spectral type bin.  In comparing Figure~\ref{fig:sp_act_sep} to Figure~\ref{fig:act_spt}, the activity
fraction across the spectral types follows the same trend for each
separation bin as it did for the total population: low activity
fractions in early spectral types and increasing activity fractions
toward the later spectral types.  We find that the closer
separated pairs are more active than the wider separation pairs, while
the widest pairs are still more active (and likely to be active) than
their field counterparts.  We also report an interesting trend that is seen at
the later spectral types, where the active fractions for three different separation
populations, as well as the field population appear to converge.  Together with other results in this
paper, we suggest that the mechanism driving magnetic fields in
late-type dMs becomes decreasingly sensitive to the presence of a
close companion, which we have interpreted as an insensitivity to
rotation.  However, Figure~\ref{fig:lhalbol_sep} shows that while every spectral type bin shows signs of decreasing activity with
increased separation, a linear least-squares fit is only at most 50\% significant over a flat horizontal line drawn through the data for each spectral type.   Figure~\ref{fig:sp_act_sep} shows that the activity state is highly dependent on the separation of the WD$+$dM binary system while Figure~\ref{fig:lhalbol_sep} shows that evidence for a correlation between activity strength and separation of the WD$+$dM binary system is inconclusive.  As a result of our study and compilation of the WD$+$dM sample we have
found a number of objects that have shown convincing signs of being in
a very tight binary system.  We have identified 37 objects in Table~\ref{tab:wddmrv}
that we feel may be valuable to the astronomical community in studying
dynamics of close binary systems, mass accretion, eclipsing
binaries, and pre-cataclysmic variables.

\section{Summary}
Using close WD$+$dM pairs, we have investigated the 
magnetic activity properties of M dwarfs in WD$+$dM pairs.  We
used SDSS DR8 spectra of unresolved binaries and developed an
iterative procedure to separate the two spectral components in 1756
high S/N WD$+$dM pairs.  We used
the H$\alpha$ line as an indicator of magnetic activity and both
absolute Galactic height (pc) and WD cooling models to examine the
activity properties as a function of age.  Both the activity state and the activity strength (via
\lha~) as a function of separation were investigated using approximate
binary separations found from the dM and WD component radial
velocities.  Below we summarize the findings of this paper.

\begin{enumerate}
\item We built upon previous close WD$+$dM investigations
  (\citealt{Raymond2003};~S06;~\citealt{Rebassa2010}) and compiled a sample of 1756 high
  S/N close WD$+$dM pairs.
\item We calculated numerous parameters for our sample including: dM spectral type, WD
effective temperature, WD surface gravity, WD and dM
relative radial velocities (km s$^{-1}$), the system
radial velocity through space (km s$^{-1}$), galactic height above
the plane (pc), approximate distance to the system (pc), approximate
binary separation (AU), period of binary (days), activity (as traced by H$\alpha$),
and \lha, and WD ages from cooling models.

\item We confirmed that paired dMs are more active than field dwarfs for
  early dM spectral types but then become comparable in their
  activity fractions at later spectral types.  This gives further
  evidence to the notion that the mechanism that generates magnetic fields in late-type dM spectra may be
  different than the mechanism in early-type stars, and that it may be
  relatively unaffected by a close companion.

\item The activity lifetimes of early-type paired dMs is longer than
  field dMs, but still shows sign of decreasing with increasing height,
  implying a finite lifetime.  However, late-type paired dMs have
  lifetimes consistent with the lifetimes found in isolated dMs.  This is reiterated when using WD cooling ages and finding
  that early-type active dMs in WD$+$dM pairs are on average younger and
  part of a distinct population in comparison to 
  inactive early-type dMs in WD$+$dM pairs.  In contrast, the late-type
  active dMs in WD$+$dM pairs span a broader range in age and are
  indistinct from the late-type inactive dMs in WD$+$dM pairs.

\item We found that the median activity strength in WD$+$dM pairs is
  larger across every spectral type when comparing to the field dM
  population.  When analyzing activity strength in the proper Galactic
  context we find that spectral types M0-M2 \lha~remains constant for
  both populations; M3-M4 field dMs exhibit a decreasing \lha~with
  increasing galactic height while WD$+$dMs do not exhibit a decreasing
  trend and are still elevated in \lha~in comparison with the field
  population; M5 field dMs again exhibit a decreasing \lha~with
  Galactic height, again not seen in WD$+$dMs, M6+M7 field dMs show a
  significantly lower \lha~than in earlier spectral types while M6+M7
  did not decrease in \lha~as significantly as the field population.

\item The closer the separation between WD$+$dM (and presumably any X+dM
  system) the more likely the dM is to be active.

\item The activity strength in WD$+$dMs shows no conclusive evidence for being
  dependent upon the proximity of the binary pair.

\item We present 37 candidate very close pairs exhibiting significant radial
  velocity changes on hour timescales to which we hope will be useful
  for the community for follow up studies.
\end{enumerate}

In investigating the activity properties of M dwarfs with close WD companions we have
found some interesting trends occurring near the transition at which
dMs become fully convective.  Spectral types ranging from M0-M7 have
been shown to exhibit higher activity fractions than the field dM
population (Figure~\ref{fig:act_spt}).  However, a possible convergence between the
WD$+$dM and field dM population is seen at later spectral types.  When
put in the proper Galactic context, we show that early-type WD$+$dMs
still exhibit higher activity fractions as
well as extended (albeit finite) activity lifetimes when compared to
the field population.  Beginning at spectral type of M5, we find that
the activity behavior in WD$+$dM becomes very similar to that of the
field population; both in activity fractions and lifetimes.  This
change conveniently occurs around the proposed $\sim$M4 transition to
fully convective envelopes.  We argue that the primary difference
between the WD$+$dM and field population is that the rotation of the dM
is being sped up by the presence of a close WD companion; most likely
through the destruction disk and thus reducing magnetic braking in the
dM and strengthening rotation \cite{Meibom2007}.  The decrease in
activity with the supposed increased rotation in late-type WD$+$dM
implies that the magnetic field generation mechanism is not as
sensitive to
rotation as in early-type M dwarfs.  Almost in contradiction, we find that the activity strength does seem to increase in the presence of a WD
companion in late-type WD$+$dM.  We have also shown that \lha~increases
with the proximity of the binary companion.  With our results we
suggest that rotation in late-type dMs has a bimodal effect; rotation
may play less of a role in establishing magnetic activity but does have an
effect on the strength of the magnetic activity.

There are, however, alternative solutions. The first one being that
rotation is important for activity and the proximity of a close binary companion is not a reliable indicator
of the dM being ``sped'' up in comparison to a field dM.  Another
solution is that at later spectral types there is a lower rotation 
threshold that sets the activity and most of the late-type stars in both the
populations are above that threshold.  This solution is put into
question by the study by \cite{WestBasri2009}, which presented
field M7 dMs with significant rotation but no measured activity.  
Also, H$\alpha$ emission can be induced on the dM by the irradiation of a hot WD companion in close binary systems \citep[e.g.][]{Tappert2011,Parsons2012}. However, the effects of the irradiation only becomes significant for systems with WD $T_{\textrm{eff}} > 45,000$K \citep{Tappert2011}. Only 78 (4\%) of our sample have $T_{\textrm{eff}} \ge$ 45,000 K with the majority having $T_{\textrm{eff}}$ between $~10000$K and $~25000$K. Hence, H$\alpha$ emission in dM induced by WD irradiation is expected to be negligible for 96\% of our sample; we expect it to have little effect on the results presented in this paper.
 
In the future, we hope to increase the sample size of WD$+$dM pairs with
spectral types $\ge$ M6 and extending this analysis to lower masses in
addition to improving the overall statistics.  Despite the current dearth of
late-type dMs in WD$+$dMs companions, we
have added to previous studies that find likely differences in the
dynamo mechanism across the fully convective boundary.  We believe
that close WD$+$dM pairs provide some unique insight into the
behavior of magnetic fields across the fully-convective stellar mass
regime and will aid in constraining the models for generation and maintaining
magnetic fields in fully convective M dwarfs.

\acknowledgments
The authors would first and foremost like to thank the anonymous referee for his or her thoughtful review and comments, which greatly improved the quality of the manuscript.  We thank also John Bochanski for useful discussions and suggestions throughout the development of this paper, as well as  Ren\'{e} Heller, Ada Nebot Gomez-Moran, Boris Gaensicke, and Alicia Aarnio for helpful suggestions and comments. In addition, we thank Bertie Wright, Jessica Stellman, Kyle
Schluns, David Jones, and
Antonia Savcheva in making note of any possible WD$+$dM pairs during
their respective studies.  We thank Ralf Napiwotzki for providing us with his fitting code fitsb2 and for helpful discussions on the Balmer line fitting and D. Koester for providing us with his white dwarf models.

AAW+DPM acknowledge the support of the NASA/GALEX grant program under
Cooperative Agreement No. NNX10AM62G issued through the NASA Shared
Services Center. This study is based on observations made with the
NASA Galaxy Evolution Explorer. GALEX is operated for NASA by the
California Institute of Technology under NASA contract NAS5-98034.
AAW+DPM also acknowledge funding from the NSF grant AST--1109273 (P.I. A. West).  A.G. 
acknowledges support from the Spanish MICINN grant AYA2009-06934. S.C. 
acknowledges support from the 7th European Community Framework Programme 
through a Marie Curie Intra-European Fellowship. S.D. was funded by NSF grant AST-0909463 (PI: K. Stassun).  In addition, M. F. acknowledges funding from KINSC Summer Research Fund Account \#3039-050.

    Funding for the SDSS and SDSS-II has been provided by the Alfred P. Sloan Foundation, the Participating Institutions, the National Science Foundation, the U.S. Department of Energy, the National Aeronautics and Space Administration, the Japanese Monbukagakusho, the Max Planck Society, and the Higher Education Funding Council for England. The SDSS Web Site is http://www.sdss.org/.

   The SDSS is managed by the Astrophysical Research Consortium for
    the Participating Institutions. The Participating Institutions are
    the American Museum of Natural History, Astrophysical Institute
    Potsdam, University of Basel, University of Cambridge, Case
    Western Reserve University, University of Chicago, Drexel
    University, Fermilab, the Institute for Advanced Study, the Japan
    Participation Group, Johns Hopkins University, the Joint Institute
    for Nuclear Astrophysics, the Kavli Institute for Particle
    Astrophysics and Cosmology, the Korean Scientist Group, the
    Chinese Academy of Sciences (LAMOST), Los Alamos National
    Laboratory, the Max-Planck-Institute for Astronomy (MPIA), the
    Max-Planck-Institute for Astrophysics (MPA), New Mexico State
    University, Ohio State University, University of Pittsburgh,
    University of Portsmouth, Princeton University, the United States
    Naval Observatory, and the University of Washington.

This publication makes use of data products from the Two Micron All Sky Survey, which is a joint project of the University of Massachusetts and the Infrared Processing and Analysis Center/California Institute of Technology, funded by the National Aeronautics and Space Administration and the National Science Foundation.  Also, this work is based in part on data obtained as part of the UKIRT
Infrared Deep Sky Survey.  We acknowledge the contribution of the JHU Sloan Digital Sky Survey
group to the development of this site. Many of its features were
inspired by the look and feel of the SkyServer. A special thanks goes
to Tamas Budavari for his help with the SDSS-GALEX matching and to Wil
O'Mullane for his help with the CASJobs site setup and configuration.
Finally, we also acknowledge Randy Thompson (MAST) for providing IDL IUEDAC
routines and Mark Siebert for providing IDL routines to generate tile
JPEG images. 

\bibliography{ms}
\bibliographystyle{apj}

\end{document}